\documentclass[final,3p,times]{elsarticle}

\usepackage{amssymb}
\usepackage{amsmath}

\usepackage{xcolor}
\usepackage{url}

\journal{Aerospace Science and Technology}

\begin{document}

\begin{frontmatter}



\title{Cost-effective multi-fidelity strategy for the optimization of high-Reynolds number turbine flows guided by LES}

\author[label1]{Camille Matar} 
\author[label2]{Paola Cinnella} 
\author[label1]{Xavier Gloerfelt} 

\affiliation[label1]{organization={Arts et Métiers ParisTech},addressline={151 bvd de l'Hôpital}, 
            city={Paris},
            postcode={75013}, 
            state={Ile-de-France},
            country={France}}
\affiliation[label2]{organization={Sorbonne University},addressline={6 place Jussieu}, 
            city={Paris},
            postcode={75006}, 
            state={Ile-de-France},
            country={France}}

\begin{abstract}
A cost-effective multi-objective shape optimization strategy is proposed for high-Reynolds number flows involving complex phenomena such as boundary layer transition, shock-wave interactions, and turbulent wakes. These processes are poorly captured by Reynolds-Averaged Navier--Stokes (RANS) models, necessitating higher-fidelity approaches like Large Eddy Simulation (LES). However, LES is computationally prohibitive at high Reynolds numbers, making its direct use in optimization impractical. To address this, we introduce a low-dimensional design space representation using Singular Value Decomposition (SVD) and construct a multi-fidelity co-Kriging (MFK) surrogate model combining wall-resolved LES (WRLES) and RANS. Adaptive infill criteria are employed to strategically enrich the surrogate model within a limited computational budget (fewer than 10 LES samples). The methodology is applied to optimize a supersonic turbine vane for Organic Rankine Cycles (ORC), operating at Reynolds numbers of $\sim10^6$. While RANS-LES correlation weakens near the optimal region, the MFK model outperforms single-fidelity Kriging (SFK) trained on the same LES data, effectively leveraging both abundant low-fidelity and scarce high-fidelity data. RANS accurately predicts global objective function trends but fails to resolve key flow features, whereas the MFK model captures fine-detail geometry trends from LES. Loss analysis reveals that LES is essential for identifying performance-detrimental mechanisms, while RANS-only optimization yields sub-optimal designs.

\end{abstract}




\end{frontmatter}


\section{Introduction}
Advanced aerodynamic shape optimization tools have become increasingly capable of producing efficient turbine vane geometries for many applications ranging from aerospace transportation to energy production \cite{Lavimi2024}, and have recently been applied to the design of Organic Rankine Cycle (ORC) power plant components \cite{jubori2017,Witanowski2020}, and specifically ORC turbine expanders. ORCs are used for power generation from low- to moderate-temperature heat sources, e.g. in waste heat recovery applications and solar thermal power plants. 
Both have attracted interest for aerospace applications \cite{Krempus2023,Pateropoulos2021}, in addition to their traditional application to terrestrial power generation (see \cite{Colonna2015} for a review).
For single-objective optimization problems, shape design methods range from complex adjoint flow solvers to simpler gradient-based and gradient-free algorithms, of which comprehensive reviews are given in \cite{Martins2020,Haji2021,Lavimi2024}.
In particular, the adjoint method has proven to be a reliable and efficient approach to turbomachinery design, for instance in \cite{Vitale2020} to concurrently generate optimal blade shapes of three rows of an axial turbine for ORC application.
However, adjoint optimization is a local optimization method, and as such it is not straightforwardly applicable in multi-modal global optimization problems, problems with complex nonlinear constraints, and multi-objective problems \cite{Forrester2008,forrester2009recent}. Most importantly, adjoint methods struggle when applied to systems involving chaotic unsteady phenomena: for this reason, their coupling with so-called scale-resolving simulations of turbulent flows, i.e. simulations resolving part of the chaotic turbulent scales, is an open research field \cite{karbasian2022sensitivity,belme2024adaptation,fidkowski2025practical}.

In the case of complex global multi-objective and multi-modal optimization problems, where the simultaneous minimization of competing quantities is performed, particle swarm \cite{Duan2016,Bahrani2015}, simulated annealing \cite{Tiow2002,Leylek2010}, or a subclass of evolutionary algorithms, such as Genetic Algorithms (GAs), is typically employed.
Specifically, the latter have been extensively applied to turbomachinery configurations \cite{Benini2001,Derakhsan2010,Li1997,Oksuz2002,Trigg1999,Dennis2001}, and more specifically to ORC expanders \cite{jubori2017,Witanowski2020,Sarmiento2021,Congedo2011a,Bufi2016,Serafino2023,Hu2021,Persico2018a,Yu2023}.

In the various works cited above, the objective functions to be minimized consist in the losses accumulated in the flow across the turbine cascade.
These losses originate from several complex flow phenomena, such as the blade boundary layers evolution (e.g. from laminar to turbulent), or the irreversible mixing taking place inside the wake behind the trailing edge (TE) \cite{Xu1988,Denton1993}.
Because of the inherent turbine complex geometries and the associated high cost of simulations, the Reynolds-Averaged Navier-Stokes (RANS) approach is the state-of-the-art model used to estimate the various loss figures with an amenable computational budget.
However, RANS models fail to accurately reproduce several key flow mechanisms that play a major role in the generation of turbine losses, such as the total momentum deficit due to boundary layers or the base pressure behind the TE \cite{Sandberg2022}, and the use of more reliable, high-fidelity computational models is urgently needed to manage the transition to a new generation of low-carbon machines \cite{Hammond2022,Tyacke2019}.
This is also true for ORC turbines, especially at operating conditions where laminar/turbulent transition, shock wave/boundary layer interactions and wake unsteadiness play an important role, as revealed by recent studies relying on high-fidelity simulations \cite{Giauque2023,gloerfelt2024high,gloerfelt2025roughness,bienner2025investigation}.
Unfortunately, the cost of higher-fidelity, turbulent-scale-resolving methods such as Large Eddy Simulation (LES) or hybrid RANS/LES is too prohibitive to be used for many-query applications such as shape optimizations. While a few LES-based optimizations have been reported very recently \cite{zhang2024large,hamedi2025far}, these remain limited to relatively low-Reynolds-number flows, for which multiple queries of the LES solver, although costly, are still feasible.
Such approaches are not expected to scale to the very high Reynolds numbers encountered in most engineering applications, and more specifically in ORC turbines. This has essentially limited aerodynamic design in general, and specifically ORC turbine design, to 2D or 3D RANS simulations only \cite{Pinto2017}. 
As a result, the designs are optimal under the RANS equations at best, with potential misrepresentation of the actual performance.
This work aims at including high-fidelity (HF) simulations within the optimization loop at an accessible cost to guide the procedure towards realistically better performing designs.

Because of the long run times associated with computational fluid dynamics solvers, surrogate (or response-surface-based) optimization methods are used to speed up the search
processes, avoiding too many calls to the costly simulators to evaluate the problem cost functions. 
A wide variety of surrogate models exist in the literature \cite{Li2017}, and much effort has been put in their implementation in aerodynamic design.
Among the available strategies, Kriging models have been successfully used in various applications (including ORC turbine vane design) and have shown good ability to provide accurate estimates of the better performing designs \cite{Rodriguez-Fernandez2015}.
Both single- and multi-objective global optimizations of ORC blade shape have been performed using Kriging surrogates \cite{Persico2019,Sarmiento2021}, including robust optimization under uncertainty \cite{Congedo2011a,Bufi2016,Serafino2023}.
An attractive characteristic of Kriging methods is that they offer a natural framework for fusing data of different levels of fidelity by adopting a so-called multi-fidelity (MF) formulation \cite{Gratiet2014}.
MF Kriging models allow inclusion of HF data within the design loop, making the optimization procedure capable of producing more realistic designs while maintaining an affordable turnaround cost for the optimization process.
The topic of MF models is currently under intense research, with very recent developments of MF Artificial Neural Networks (ANNs) \cite{Meng2020}, applied to single-objective aerofoil shape optimization \cite{Zhang2021,Wu2024}.
A thorough review of MF methods for optimization is provided in \cite{Li2024}.
While several approaches exist to build multifidelity surrogates, most of them assume some form of correlation between the low fidelity and high fidelity model. While nonlinear correlation models have also been proposed, they generally require more high-fidelity data, which can be too expensive or unfeasible in the case of very expensive high-fidelity models such as those involving scale-resolving turbulent flow simulations (see \cite{brevault2020overview} for an overview).

In the effort of seeking LES-quality optimal ORC turbine vane designs at an affordable cost, we investigate here a MF Kriging approach where only very few, well-chosen designs are simulated with the costly HF LES approach, while the rest of the design space is explored by the low-fidelity (LF) RANS model.
For that purpose, we follow the autoregressive co-Kriging procedure described in \cite{Gratiet2014} to generate a nested MF co-Kriging (MFK) model, based on the assumption that the RANS and LES predictions are sufficiently well correlated, and a linear relationship between both fidelity levels can be established. 
We also devise a very few-shots strategy for correcting the LF-based Kriging with LES of designs with high probability of improving the optimization criteria.
A series of tests is conducted to identify the best strategy for improving a LF surrogate using very few HF samples, and the validity of the underlying correlation hypothesis and its effect on the resulting optimum is assessed.
%
Finally, we carry out a detailed flow analysis to show the impact of turbulence modeling fidelity on the selection of optimal blade shapes.
%



\section{Case study and numerical flow models}
\subsection{Case study}
ORCs are Rankine cycles for low-temperature applications and use organic vapors as their working fluid.
These compounds are generally characterized by high molecular complexity and molecular weight, large heat capacity, and low
critical temperature compared to steam.
In typical ORC operating conditions the dense organic vapor often exhibits non-ideal gas dynamic behaviors \cite{Guardone2024} induced by its complex thermodynamic response.
In an ORC a fluid is compressed, vaporized and then expanded to produce a power output.
The expansion is generally realized in a turbine for ORCs of medium/large size.
ORC turbines are typically small devices subject to very high pressure ratios, which leads to transonic or even supersonic flow regimes at the exit of the primary stator vane.
This results in the formation of shock waves disrupting the blade boundary layers and possibly triggering transition to turbulence.
Furthermore, the typically high density of dense vapors causes the Reynolds number to rapidly increase on the blades, exceeding $10^6$.
The gas dynamic behavior of dense organic vapors is well described by the fundamental derivative of gas dynamics \cite{Thompson1971}
\begin{equation}\label{fund deriv}
\Gamma := \frac{v^3}{2c^2}\frac{\partial^2 P}{\partial v^2}\bigg\rvert_s = 1+\frac{\rho}{c}\frac{\partial c}{\partial \rho}\bigg\rvert_s
\end{equation}%
which indicates how the speed of sound $c=\sqrt{(\partial P/\partial\rho)_s}$ varies with density $\rho=1/v$ across isentropic processes, where $v$ is the specific volume, $P$ is the pressure  and $s$ is the entropy.
For a thermally and calorically perfect gas, $\Gamma=(\gamma+1)/2>1$ and constant, $\gamma=c_p/c_v$ being the specific heat ratio (isentropic exponent), with $c_p$ and $c_v$ the isobaric and isochoric specific heats, respectively.
On the other hand, organic vapors of sufficient molecular complexity possess a region where $\Gamma<1$, delimiting the dense gas region, which implies that the rate of change of speed of sound with respect to density is negative across isentropic transformations.
Therefore, $c$ drops through compressions and grows through expansions in an opposite fashion to an ideal gas.
Finally, the mean of identifying non-ideal thermodynamic conditions is to re-write the ideal gas equation with the compressibility factor Z
\begin{equation}\label{compressibility factor}
    P = Z\rho RT
\end{equation}%
where R is the gas constant and $Z=Z(\rho,T)$ is a function of two thermodynamic quantities.
Similarly, dense gases exhibit values of $Z<1$ as they approach the liquid-vapor saturation line, and real-gas effects become prominent.
While thermodynamic conditions leading to severe departure from ideal gas dynamics exist, whereby the Mach number variation with density is no longer monotonic \cite{Cramer1991}, the present study focuses on regimes where the dense gas behaves as a dilute gas, where $\Gamma$ and $Z$ are close to unity and where the organic vapor deviates only mildly from the behavior of an ideal gas.

%

In the following, we focus on the optimal design of the cross section of a supersonic ORC turbine stator.
The latter is modeled as a linear cascade, which is a reasonable approximation since ORC blades are typically untapered and untwisted. 
All 3D effects due to the side boundary layers are neglected in the study, and the flow around the blades is assumed to be statistically two-dimensional, so that only the flow in a cross-sectional plane is simulated with the RANS equations, and an extruded spanwise portion of the domain is simulated with LES.

We consider a baseline ORC vane geometry  initially designed by \cite{Baumgartner2020} by using a modified method of Characteristics, which assumes inviscid potential flow throughout the supersonic portion of the turbine vane to account for the non-ideal behavior of the gas during the supersonic expansion. 
The geometry of the subsonic portion and of the rounded trailing edge is based on existing best practices. 
The baseline geometry is visible in Figure \ref{LES baseline grid}.a. 
Since the blade design is based on an inviscid flow assumption, we do not expect it to be optimal in the presence of viscous effects due to the development of the boundary layers and wake. 
Furthermore, at the present supersonic operating conditions, a system of shock waves (not accounted for by the potential flow model) is generated at the blade trailing edge, which interacts with the surrounding boundary layers and wakes.
\begin{table}[t]
	\centering
	\begin{tabular}{c|c|c|c|c|c|c|c|c|c}
		\textbf{Fluid} & $\mathcal{M}$ & $T_c$ & $P_c$ & $\rho_c$ & $R$ & $c_{v,\infty}$ & $n$ & $\bar{\omega}$ & $\bar{\xi}$ \\
		& [g/mol] & [K] & [MPa] & [kg/m$^3$] & [-] & [J/kg/K] & [-] & [-] & [Debye] \\
		\hline
		R134a    & 102.03 & 374.21 & 4.06 & 511.90 & 81.489 & 1328.27 & 0.480 & 0.327 & 2.058 
	\end{tabular}
	\caption{Refrigerant R134a thermodynamic properties, where $\mathcal{M}$ is the molecular weight, $T_c$, $P_c$ and $\rho_c$ are the critical temperature, pressure and density, $R$ the gas constant, $c_{v,\infty}$ the dilute gas limit isochoric heat capacity, $n$ the power-law exponent for $c_{v,\infty}$, $\bar{\omega}$ the acentric factor and $\bar{\xi}$ the dipole moment.}
	\label{tab: fluid properties}
\end{table}%

The blade is operated using the refrigerant R134a, often adopted in ORC cycles, as the working fluid. 
The main thermophysical properties of this gas are reported in Table \ref{tab: fluid properties}.
The inlet and outlet thermodynamic conditions, along with the target outflow angle $\beta_2$ and imposed pressure ratio $\Pi=P_{0,1}/P_2$ of the gas ($P_{0,1}$ and $P_2$ being the inlet stagnation and exit static pressures, respectively) are summarized in Table \ref{tab: Operating conditions}.
At the chosen operating conditions, the Reynolds number based on the blade axial chord $C=10$~mm and the isentropic exit velocity is $Re_C=1.3\times10^6$, and the target isentropic Mach number downstream of the blade (i.e. the Mach number that would be reached for an ideal flow with no losses) is $M_{2,is}=1.70$.
The fundamental derivative of gas dynamics $\Gamma<1$ indicates that the thermodynamic conditions of the R134a are located in the dense gas region (see Section \ref{Governing equations and models}). This yields opposite variations of the speed of sound across isentropic transformations compared to a perfect gas \cite{Thompson1971}.
However, the compressibility factor $Z=0.90$ remains close to one, so that only mild deviations from the ideal gas behavior should be expected.
Finally, we assume that the incoming flow has a negligible turbulence intensity, i.e. the flow is assumed to be laminar at the turbine inlet.

\begin{table}
    \begin{center}
        \caption{Vane operating conditions with R134a}\label{tab: Operating conditions}
        \begin{tabular}{c|c|c|c|c|c|c|c|c}
        $Re_C$ & $M_{2,is}$ & $\Pi$ & $P_{0,1}$[bar] & $T_{0,1}$[K] & $\beta_2$[\textdegree] & $\gamma$  & $\Gamma$ & $Z$\\
        \hline
        $1.3\times10^6$ & 1.70 & 4.38 & 4.83 & 293 & 75 & 1.10 & 0.95 & 0.90 \\
        \end{tabular}
    \end{center}
\end{table}%

\subsection{Numerical models}\label{Governing equations and models}
Numerical simulations of the turbine flow are carried out using {\sc Musicaa}, a well-validated,
high-performance in-house numerical solver for real-gas flows (see \cite{Bienner2024a} for a more detailed description and assessment of the flow solver).
The flow is modeled by the compressible Navier--Stokes equations for a Newtonian fluid, supplemented with suitable thermophysical models to account for the non-ideal gas behavior.
In the present simulations, the organic vapor R134a is modeled by the Peng-Robinson-Stryjek-Vera equation of state \cite{Stryjek1986}.
Variations of the viscosity $\mu$ and thermal conductivity $\kappa$ with temperature and density are modeled by the Chung-Lee-Starling model \cite{Chung1988}. 
More details on the implementation and the validation of real-gas models for an organic vapor with comparable molecular complexity are given in \cite{Gloerfelt23b}.

%

The governing equations are approximated with high-order finite differences implemented on multiblock structured grids by means of coordinate transforms (further details can be found in \cite{Bienner2024a}).
{\sc Musicaa} allows for both RANS and LES within the same flow solver.
The LES simulations resolve the largest turbulence scales, while the effect of unresolved subgrid-scale motions is taken into account implicitly through the
explicit filter (used as part of the numerical discretization) that removes 
subfilter scales and provides a selective regularization. This implicit modeling strategy has been
shown to be effective \cite{Gloerfelt2019a,Gloerfelt23b} and avoids the computational overhead introduced
by the explicit subgrid-scale models.
Time integration is performed with a 4-stage low-storage Runge-Kutta algorithm, to which an Implicit Residual Smoothing (IRS) operator is added, enabling CFL$\approx5$ in the present LES.
RANS simulations make use of the Spalart-Allmaras one-equation model, following the implementation of \cite{Crivellini2019} to improve numerical robustness, and neglect the tripping source term such that the boundary layer is fully turbulent.
In this case, a local time-stepping approach is employed.



\subsection{Computational mesh and mesh morphing strategy}\label{Numerical mesh}
\begin{figure}[t]
    \centering
    \includegraphics[width=0.49\textwidth]{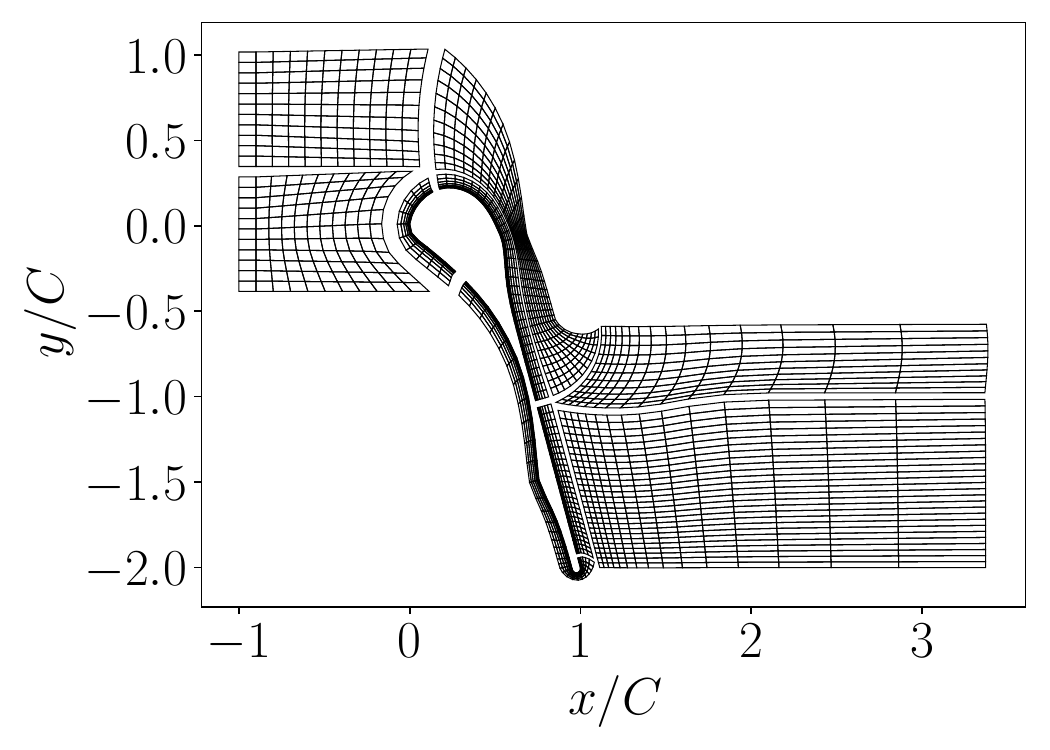}
    \hfill
    \includegraphics[width=0.49\textwidth]{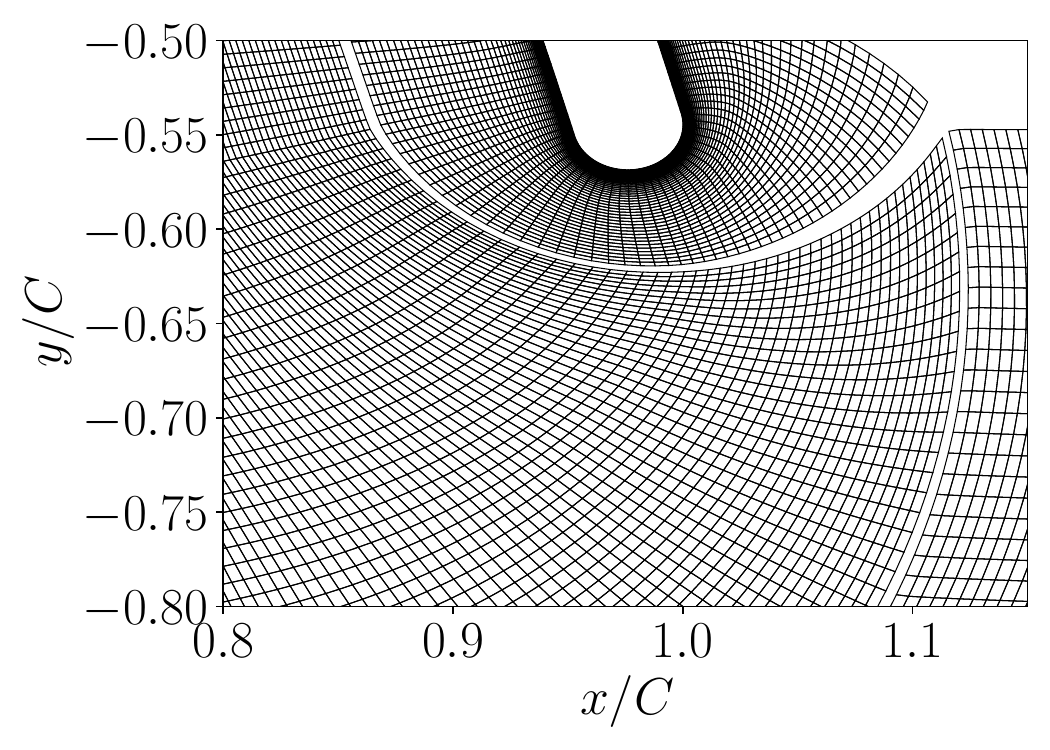}\\
    \makebox[0.49\textwidth][c]{a) Global view.}
    \makebox[0.49\textwidth][c]{b) Wake region.}
    \caption{LES grid around the baseline design. a) Global view (one in 25 points), b) details of the near wake region (one in 5 lines).}
    \label{LES baseline grid}
\end{figure}

The computational domain is discretized using a structured mesh composed of 9 blocks, and it covers $x\in[-1C;3.5C]$ in the axial direction, where $C$ is the blade chord, and we assume flow periodicity in the pitchwise direction.
We design a grid for the LES using $2\,600$ points to discretize the blade surface with clustering around the TE. 
Views of the grid around the blade and in the near wake are provided in Figure \ref{LES baseline grid}.
We then extrude the 2D grid in the spanwise direction over 10\%$C$ and 300 planes, which results in a total of 440 million points.
We assess the achieved resolution of turbulence in \ref{LES resolution}, and show that this mesh is adequate for wall-resolved LES.
We use the same grid topology for the RANS calculations, but with larger aspect ratios.
Specifically, we design a mesh with $142\,500$ points and maintain the same blade first cell height as for the LES.
We provide a convergence study of the RANS solution with mesh refinement in \ref{RANS grid convergence}.

\begin{figure}[t]
	\centering
	\includegraphics[width=0.49\textwidth]{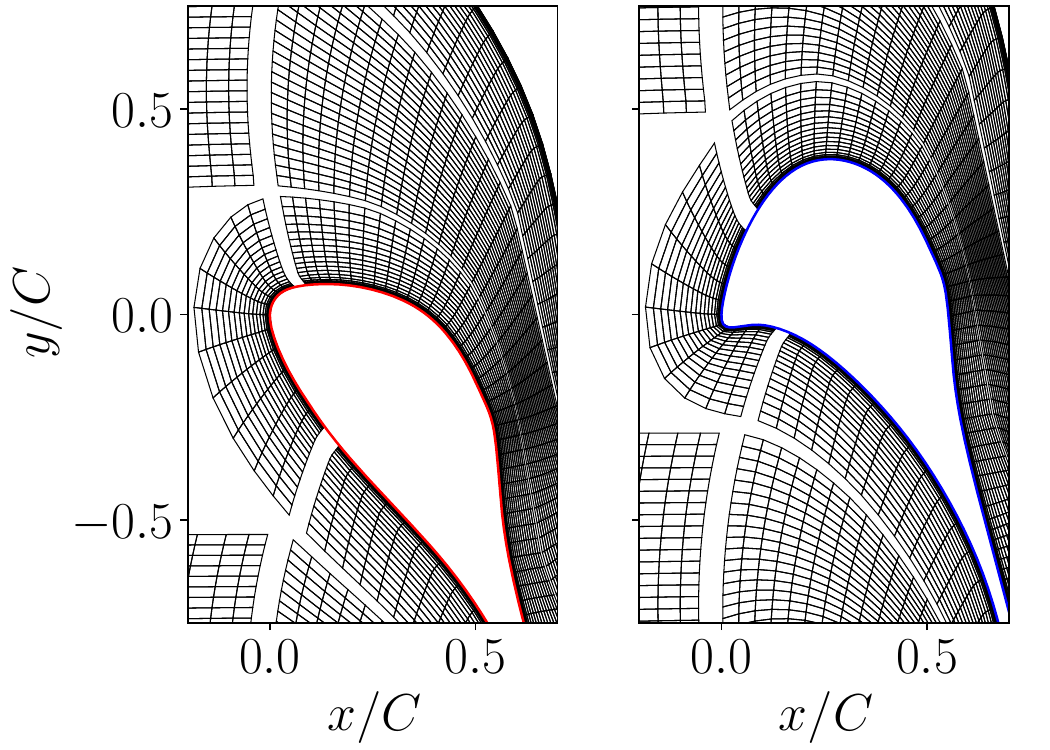}
	\centering
	\includegraphics[width=0.49\textwidth]{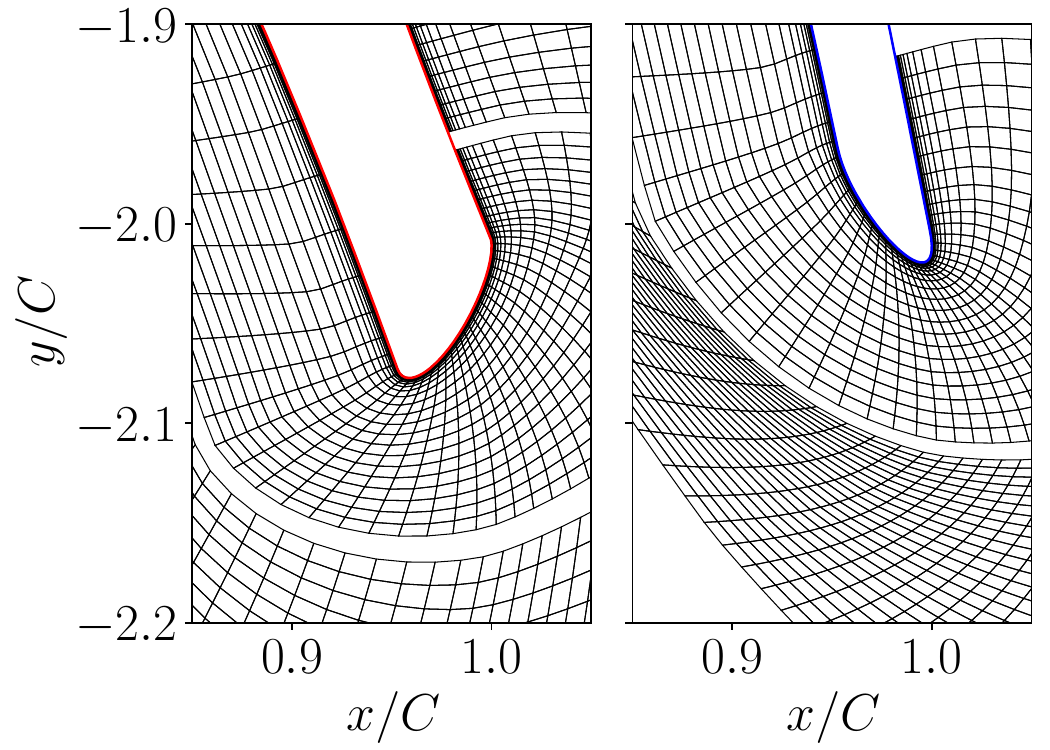}\\
	\makebox[0.49\textwidth][c]{a) Leading edge.}
	\makebox[0.49\textwidth][c]{b) Trailing edge.}
	\caption{Detailed view of the RANS mesh after extreme deformation of the baseline geometry (1 in 4 lines shown).}
	\label{fig: design min max grids}
\end{figure}

As the optimization proceeds, the mesh must be regenerated or adapted to the new blade geometries.
Because the present CFD solver is limited to structured grids, we implement a simple mesh morphing strategy based on inverse weighted interpolation of a block's boundary perturbation (due to the deformed blade shape) to its interior grid \cite{Witteveen2009}.
Specifically, we first deform the blocks surrounding the blade with the new geometry.
Then, we deform their immediate neighbors based on the new block frontiers.
Finally, we apply the same morphing strategy to the remaining ones with the exception of those containing the inlet and outlet faces.
These faces are simply fixed to ensure the physical boundary conditions behave identically for all blade designs.
We show in Figure \ref{fig: design min max grids} details of the resulting meshes around the most extreme blade deformations allowed by the present parametrization strategy (detailed in Section \ref{Blade parametrization and design variables}).
Special care was taken to ensure that mesh line orthogonality with the blade surface was only mildly affected.

\subsection{Optimization problem}
Given a search space $\Omega\subset{\mathbb R}^N$, with $N$ the number of design variables, we seek to simultaneously minimize the following set of cost functions:
\begin{equation}
    \min_{\mathbf{d}}
    \begin{cases} 
        J_1(\mathbf{d}) = \dfrac{\langle\overline{s}\rangle_{out}}{\langle\overline{s}\rangle_{in}}
         \\[2ex]
        J_2(\mathbf{d}) = \mu\left[ \left( \langle\overline{M}\rangle_{out} - \overline{M}_{out} \right)^2 \right]
    \end{cases}
\end{equation}%
under the constraint:
\begin{equation}
     \dfrac{A(\mathbf{d})}{A_\text{baseline}}\ge 1
\end{equation}
In the above, $\mathbf{d}$ denotes a particular design, identified as a vector of design variables, $J_1$ represents the increase in (time-averaged) entropy $\overline{s}$ through the turbine vane, and $J_2$ represents the spatial variance of the (time-averaged) Mach number $\overline{M}$ across the turbine outlet section.
Specifically, the subscripts $in$ and $out$ denote the inlet and outlet sections, respectively, $\langle\bullet\rangle_{in/out}$ denotes mixed-out averaging \cite{Prasad2005} along the inlet/outlet plane, respectively, and $\mu[\bullet]$ denotes an ensemble average over the blade pitch.
The constraint is based on the conservation of the initial blade cross-sectional area $A_\text{baseline}$. 

The multi-objective optimization problem is solved through the well established multi-objective Genetic Algorithm (GA) NSGA-II \cite{Deb2002}, 
which provides a Pareto front ${\cal P}\subset \Omega$ of optimal
solutions corresponding to different trade-offs between
average performance and robustness.
Specifically, the Pareto-optimal individuals $\mathbf{d}^*$ satisfy the conditions:
\begin{equation}
    J_i(\mathbf{d}^*)\le J_i(\mathbf{d})
    \quad \forall {\mathbf{d}}^*\in{\cal P}, \mathbf{d}\in\Omega, \quad \text{and} \quad i=1,2
\end{equation}
\subsection{Blade parametrization and design variables}\label{Blade parametrization and design variables}
\begin{figure}[h!]
    \includegraphics[width=0.49\textwidth]{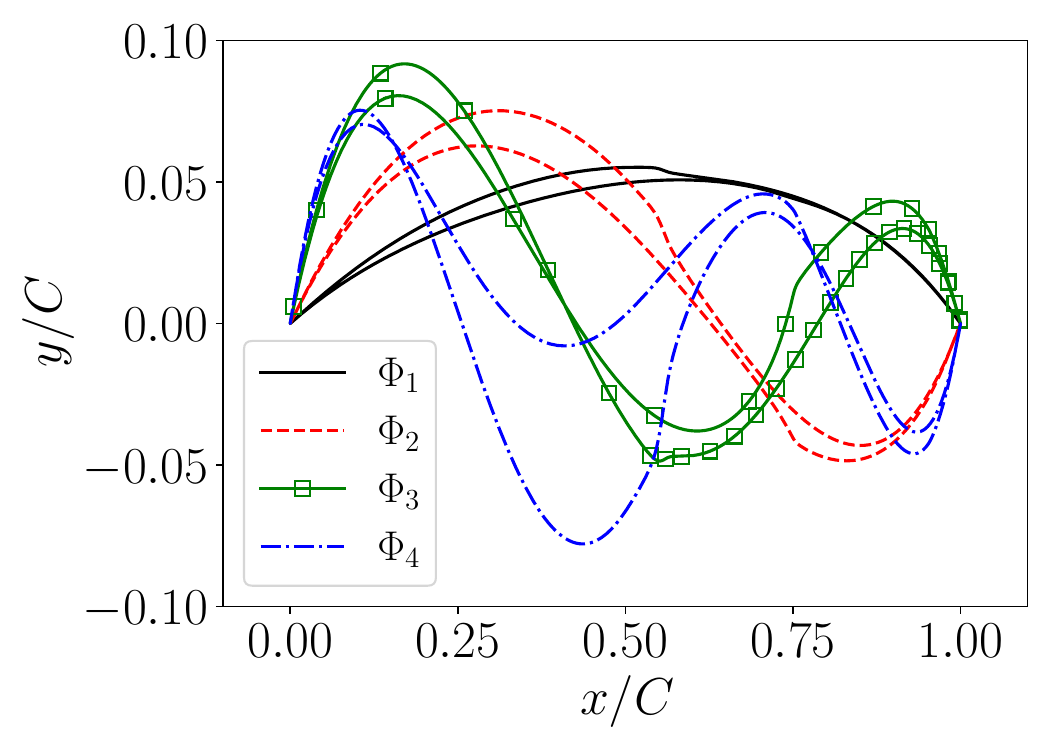}
    \quad\quad\quad
    \includegraphics[width=0.29\textwidth]{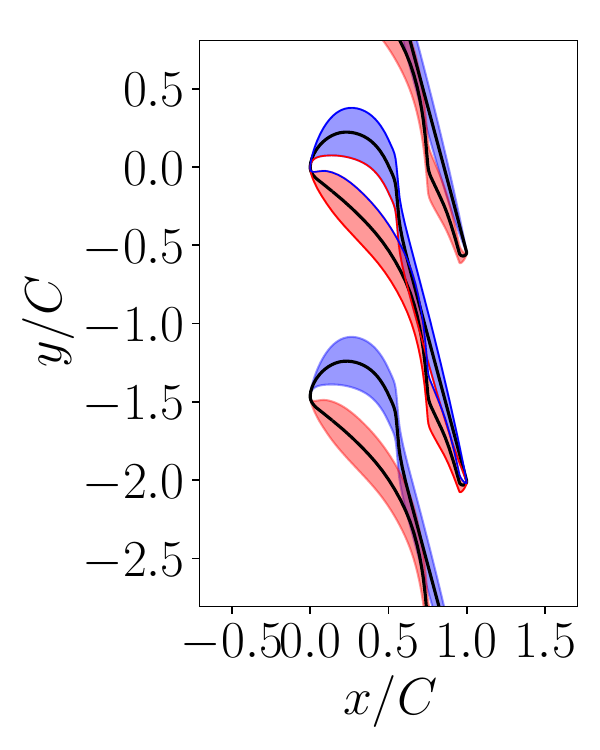}\\
    \makebox[0.49\textwidth][c]{a) Geometric modes.}
    \makebox[0.49\textwidth][c]{b) FFD deformation range.}
    \caption{a) Geometric modes extracted from the SVD of the FFD design database, where $\Phi_i$ is the $i$th mode. b) FFD deformation range.}
    \label{fig: modes}
\end{figure}

The present parametrization strategy is based on a Free Form Deformation (FFD) approach \cite{sederberg1986free}, which allows to represent complex shapes using a relatively small number of control parameters, coupled with a simple data reduction technique.
Specifically, the blade section is initially parametrized using FFD on $N_{FFD}=8$ control points, of which the only degree of freedom is their vertical displacement, and a large number of random designs is generated.
We provide an example FFD parametrization of the blade in \ref{B}.
Then, the random designs are assembled into a database matrix $\boldsymbol{D}$ over which a Singular Value Decomposition (SVD) is performed to reduce the design space dimensionality \cite{Toal2008}.
The result of the SVD writes:
\begin{equation}
    \widetilde{\boldsymbol{D}} = \overline{\boldsymbol{D}} + \widetilde{\boldsymbol{\Phi}}\widetilde{\boldsymbol{A}}^{-1}
\end{equation}%
The matrix $\overline{\boldsymbol{D}}$ has size $N_y\times N_d$, where $N_y$ is the number of blade surface points and $N_d$ the number of designs in the database, and contains the mean of the database (corresponding to the baseline geometry) along its columns.
The matrix $\boldsymbol{\Phi}$ has size $N_y\times N_d$ and contains the modes $\Phi_i$ ($i=1,...,N_d$).
The matrix $\boldsymbol{A}^{-1}$ has shape $N_d\times N_d$ and contains the modal coefficients $\alpha_i$ along its columns.
Retaining the first $t=4$ modes is deemed sufficient to represent all random designs, leading to the truncated $N_y\times N_t$ and $N_t\times N_d$ matrices $\widetilde{\boldsymbol{\Phi}}$ and $\widetilde{\boldsymbol{A}}^{-1}$, respectively.
We show the first 4 $\Phi_i$ modes in Figure \ref{fig: modes}.a.
Although these have little physical relevance, one anticipates that positive and negative modal coefficients will raise and lower the LE ($x/C\in[0;0.25]$), respectively.
Thus, the 4 modal coefficients associated to each mode constitute the parameter space.

The range of blade deformations is illustrated in Figure \ref{fig: modes}.b and results from the FFD control point displacements contained within $\pm10\%$ of the blade height.
The NSGA-II crossover rate is set to 0.9 and the mutation rate is specified individually for each modal coefficient. 
Indeed, the first mode carries over 70\% of the total modes energy and is thus considered as driving the design space exploration. 
Therefore, in the NSGA-II algorithm, a high mutation rate of 0.2 is associated to this variable, while a conservative value of 0.1 is attributed to the remaining three.

\subsection{Surrogate modeling}
\subsubsection{Multi-fidelity co-Kriging (MFK) model}
To model the objective functions response in the parameter space, we use the autoregressive MFK model formulation proposed in \cite{Gratiet2014} for $l=2$ levels of code. 
The variation of the cost functions $J_i$ with respect to the design vector $\mathbf{d}$ is represented through multi-fidelity Gaussian processes, with mean and variance given by
\begin{equation}\label{MFKrig}
    y_{h}(\mathbf{d}) = \rho_{l}(\mathbf{d})y_{l}(\mathbf{d}) + \boldsymbol{r}^T\boldsymbol{R}^{-1}(\mathbf{s}_{h} - \rho_{l}(\mathbf{d})\mathbf{s}_{l|h})
\end{equation}%
\begin{equation}\label{MFKrig_var}
    \sigma^2_{h}(\mathbf{d}) = \hat{\sigma}_{h}^2 \left[1-\boldsymbol{r}^T\boldsymbol{R}^{-1}\boldsymbol{r}  + \right. 
    \left. \left[ \boldsymbol{r}^T\boldsymbol{R}^{-1}\mathbf{s}_{l|h} - y_{l}(\mathbf{d})\left(\mathbf{s}_{l|h}^T\boldsymbol{R}^{-1}\mathbf{s}_{l|h}\right) \left[ \mathbf{s}_{l|h}^T\boldsymbol{R}^{-1}\mathbf{s}_{l|h}-y_{l}(\mathbf{d})^T \right] \right] \right]
\end{equation}%
where $y$ denotes a Quantity of Interest (QoI) (here, a cost function $J_i$), subscripts $h$ and $l$ denote the high- and low-fidelity levels, $\rho_{l}(\mathbf{d})$ is an adjustment coefficient (determined with a least-squares regression on the known data), $y_{l}(\mathbf{d})$ is the LF Kriging model prediction, $\boldsymbol{r}$ is the correlation vector between the untried point and the HF samples, $\boldsymbol{R}$ the correlation matrix of the HF samples, $\mathbf{s}_{h}$ is the vector of HF samples, $\mathbf{s}_{l|h}$ is the vector of LF samples at HF locations, and $\hat{\sigma}_h^2$ is the optimal standard deviation which maximizes the likelihood of the HF data.
In this work, we assume that $\rho_{l}(\mathbf{d})=\rho_{l}$ is a constant and that the low- and high-fidelity samples are sufficiently correlated.
The consequences of this particular choice will be addressed in Section \ref{Functional space}.
Then, a Gaussian kernel is chosen for the correlation functions.
Finally, independent MFK models are developed for each objective function.
In the following, we use the MFK implementation available through the SMT 2.0 toolkit \cite{saves2024smt}.

\subsubsection{Parallel infill strategy}\label{Infill strategy}
The quality of the estimated Pareto-optimal designs is expected to be highly dependent on the accuracy of the MFK surrogate. A high-accurate surrogate may require a much larger number of samples than the allocated computational budget.
In particular, based on a popular rule of thumb \cite{loeppky2009choosing}, at least $10\times 4$ samples are needed to build a reasonably accurate single-fidelity surrogate in the present four-dimensional search space, which largely exceeds our computational budget if LES is to be used.
However, the surrogate does not need to achieve the same accuracy everywhere.
Instead, we need the surrogate to be accurate enough to correctly identify poorly performing designs in non-optimal regions, and as accurate as possible in the region containing the Pareto-optimal designs.
This objective can be achieved by means of an adaptive infill strategy.

During the optimization, the Design of Experiments (DoE) is enriched with strategically chosen samples to reduce the Kriging models's error near the optimal region of the design space.
This particular point requires the Kriging predictor to deliver a sound estimate of the optimal design parameters, which is partly governed by the size of the DoEs at both low- and high-fidelity levels.
In the present work we consider a problem for which the LF model is much cheaper to evaluate than the HF one, such that many more samples are available compared to the HF for which the DoE is exceptionally small.
This renders the LF surrogate more likely of accurately predicting optimal infills.
We then assume that both levels of fidelity are sufficiently well correlated for the infills to lie near the optimal region for both LF and HF functions, so that the infill criteria estimated from the LF-dominated surrogate can be applied to both levels of fidelity.
The validity of this assumption will also be addressed in Section \ref{MF co-Kriging optimization}.

We propose to perform infills corresponding to three additional samples computed in parallel, and based on the LF Kriging model.
The first corresponds to the maximal multi-objective augmenting Probability of Improvement (PI) \cite{Keane2006,Zhang2018}, which for 2 objective functions reads
\begin{equation}
    \text{PI}(\mathbf{d}^*) = \text{P} \left[ y_{l,1}(\mathbf{d}^*) \leq y_{l,1}(\mathbf{d}_p) \cup y_{l,2}(\mathbf{d}^*) \leq y_{l,2}(\mathbf{d}_p) \right]
\end{equation}%
where $\mathbf{d}^*$ is an untried point, $y_{l,i}$ the LF Kriging prediction, the subscripts $1$ and $2$ designate the first and second objective functions, and the subscript $p$ denotes an individual from the Pareto set.
The optimal $\mathbf{d}^*$ to be computed with the LF function is one that maximizes $\text{PI}(\mathbf{d}^*)$ over the parameter space, and is found using a differential evolution algorithm \cite{Storn1997}.
In the view of taking full advantage of modern computer architectures, a parallel multi-objective PI criterion has been derived in the literature \cite{Zhang2021a}.
In principle, it provides optimally spaced local optima of the PI function to generate new samples covering a wide space near the Pareto front.
Nevertheless, we follow the suggestions from \cite{Liu2017} and use an alternative infilling strategy in case the PI were to fail.
Specifically, the other two infill criteria are the minimized single-objective Lower Confidence Bounds (LCB) computed for each objective function
\begin{equation}
\text{LCB}_i(\mathbf{d^*})=y_{l,i}(\mathbf{d^*}) - A\sigma_{l,i}(\mathbf{d^*}), \quad\text{where}\quad A=1
\end{equation}%
Similarly, the LCB$_i$s are minimized using differential evolution.

In addition to the LF infills, one HF sample is added to the training database every prescribed number of iterations.
Several authors have proposed multi-fidelity infill criteria, where the cost of an expensive simulation is compared to the relative improvement in objective function and reduction in surrogate model error \cite{Aye2023,Sleesongsom2020}.
If the latter balance the former, a sample is selected according to the infill criterion and computed with the HF function.
Many variants initially developed in the MFK framework are available, and currently represent an active research area \cite{Belakaria2020,He2022} (an extensive survey is provided in \cite{Haftka2016}).
Although these methods are attractive, they do not apply to situations such that only a very few shots at retraining the MFK model with HF LES samples are available.
Thus, the risk associated with a MF infill criterion failing is too prohibitive.
Furthermore, the cost ratio of an LES to a RANS computation in the present case is beyond $10^4$, preventing the potential MF infill criterion from ever selecting the HF infill to reduce the Kriging prediction error.
For these reasons, we simply choose to simulate with the HF method the PI infill chosen from the LF data, which naturally satisfies the nested property of the present MFK model.
The choice of frequency for this infill is discussed in Section \ref{MF co-Kriging optimization}, as it is related to the overall computational budget available.

Of note, once the chosen designs have been computed (with either the LF or HF functions), they are added to the current optimization generation to accelerate convergence, replacing the least-performing ones.


\section{Validation}
\subsection{RANS and LES predictions for the baseline geometry}\label{RANS and LES predictions}
\begin{figure}[h!]
    \centering
    \includegraphics[width=0.49\textwidth]{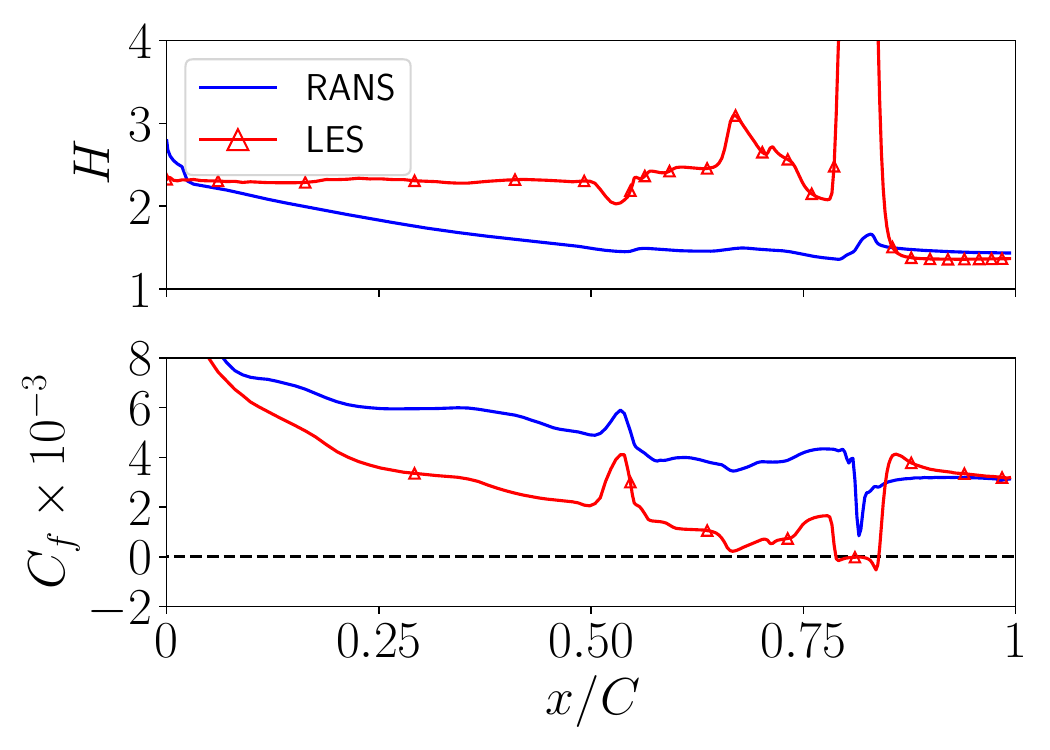}
    \hfill
    \includegraphics[width=0.49\textwidth]{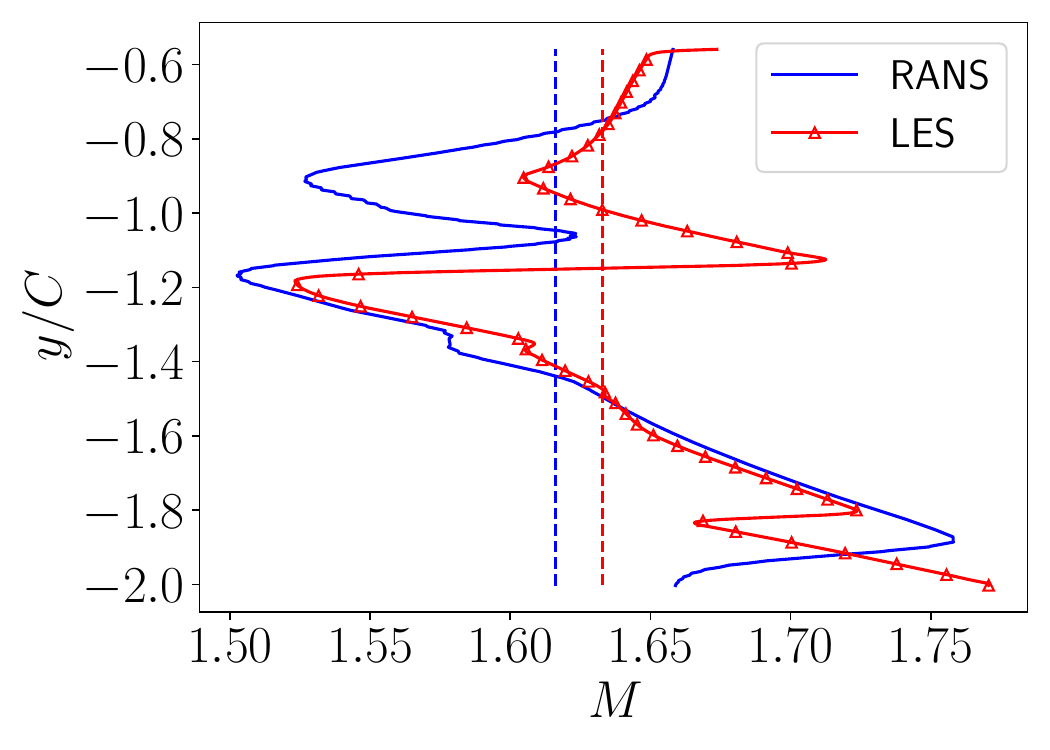}\\
    \makebox[0.49\textwidth][c]{a) Shape factor and skin friction}
    \makebox[0.49\textwidth][c]{b) Mean Mach number deviation profile}
    \caption{a) Boundary layer shape factor $H$ and friction coefficient $C_f$ on blade suction side, and b) mean Mach number $M$ deviation profiles in the wake.}
    \label{comp RANS LES}
\end{figure}

\begin{figure}[h!]
    \centering
    \includegraphics[width=0.66\textwidth]{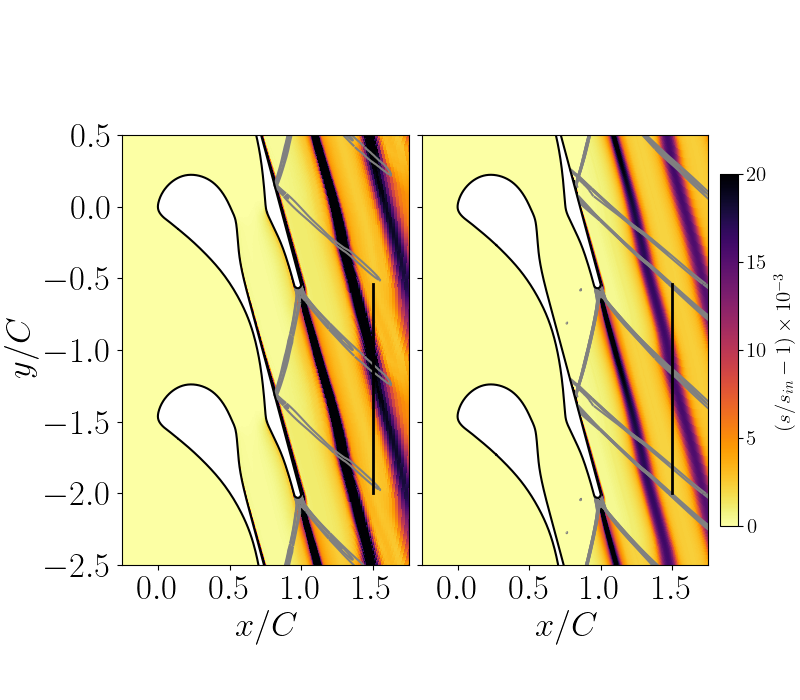}
    \caption{RANS and LES entropy fields with velocity divergence contours, and mixed-out state measurement line.}
    \label{baseline comp}
\end{figure}

To highlight the large discrepancies between LES and RANS on the baseline geometry, we show in Figure \ref{comp RANS LES}.a the boundary layer state on the blade suction side. 
The shape factor $H$ and friction coefficients $C_f$ strongly disagree on most of the blade suction side, as the RANS predicts a fully turbulent boundary layer starting from the blade leading edge (LE), as evidenced by the lower $H$ and higher $C_f$. 
Furthermore, the shockwave/boundary layer interaction (SWBLI) located at $x/C\in[0.75;0.85]$ is much wider in the case of the LES, as a plateau of negative $C_f$ exists.
As the state of the boundary layer impacts both the SWBLI and the flow at the TE, the wake and shock waves downstream of the blade can differ greatly between the two computations. 
This is shown in Figure \ref{comp RANS LES}.b, where the profile of Mach number deviation $\Delta M$ from the mixed-out state ($\langle\overline{M}\rangle=1.615;1.630$ in the RANS and LES, respectively) has been extracted along the line shown in Figure \ref{baseline comp}. 
As expected, the profiles feature large deviations and the mixed-out values differ by about $1\%$.
Finally, we give in Figure \ref{baseline comp} the mean entropy fields and velocity divergence contours obtained with both methods. 
The wake features a more pronounced entropy content in the steady RANS computation, and is slightly wider.
Furthermore, the velocity divergence contours showcase more than one reflected shock wave in the LES compared to the RANS.
Therefore, with the overestimated irreversible mixing in the wake and the disparities in Mach number distribution near the outlet, we expect large deviations in blade performance predictions between the two models.

\subsection{RANS direct and Kriging-based optimizations}
To validate the described surrogate-based optimization methodology, we first converge the optimization without a Kriging model solely using RANS on a coarse mesh, denoted LF RANS. 
With an initial population size of 30 individuals, the procedure is converged over 70 generations, resulting in a total of 2\,100 function evaluations with the flow solver (these parameters are summarized in Table \ref{tab: optim params}).
The first generation individuals are selected via a Latin Hypercube Sampling (LHS) of the parameter space.
Each LF RANS costs 4.8CPUh on 57 CPUs (31\,875 points in the mesh, 400 points discretizing the blade surface) and the optimization is run on a supercomputer, thus a total of 10\,080CPUh are required.

\begin{figure}[h!]
	\centering
	\includegraphics[width=0.5\textwidth]{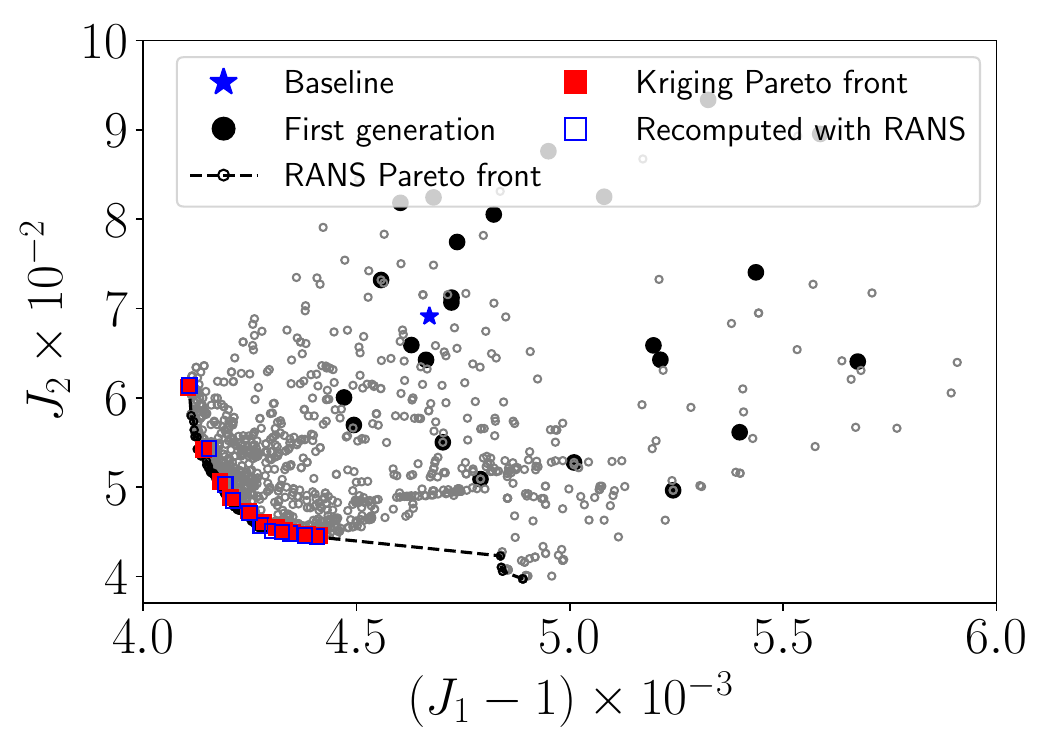}
	\caption{LF RANS direct and single-fidelity Kriging optimization functional space.}
	\label{fig: Pareto RANS LF}
\end{figure}

The resulting RANS Pareto front is shown in Figure \ref{fig: Pareto RANS LF}.
\begin{figure}[h!]
	\centering
	\includegraphics[width=0.49\textwidth]{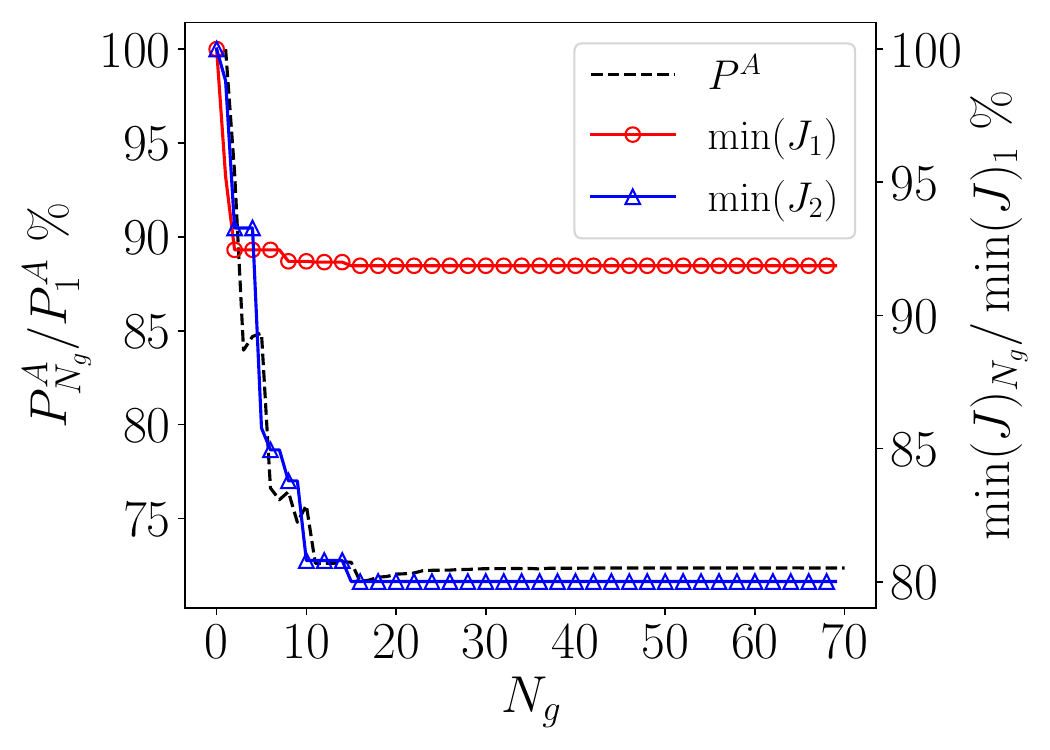}
	\centering
	\includegraphics[width=0.49\textwidth]{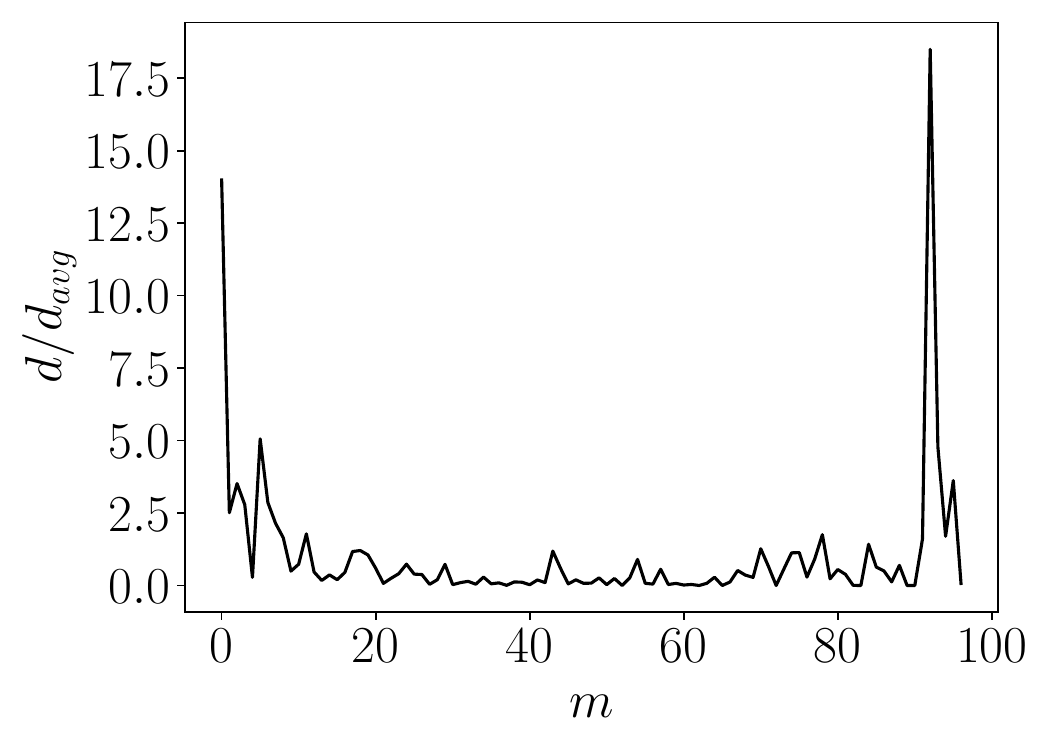}\\
	\makebox[0.49\textwidth][c]{a) Convergence metrics.}
	\makebox[0.49\textwidth][c]{b) Spacing along Pareto front.}
	\caption{a) Optimization convergence metrics: $P^A$ is the area below the Pareto front, $\min(J_i)$ are the minimum values of each objective function found during the optimization. b) Distance separating every Pareto set member from their immediate neighbor.}
	\label{fig: Pareto RANS LF Area d}
\end{figure}%
It is well distributed and features a discontinuity near the high $J_1$ and low $J_2$ limit as a result of the constraint. 
Convergence metrics are provided in Figure \ref{fig: Pareto RANS LF Area d}.a where $P^A$ is the area contained below the Pareto front and $N_g$ is the number of generations.
We estimate that the optimization has converged when $P^A$ and the minimum of each objective function cease to evolve, which in the present case occurs just after 30 generations.
Figure \ref{fig: Pareto RANS LF Area d}.b provides the Euclidean distance in the functional space separating every Pareto set member $m$ from their immediate neighbor.
This metric evaluates the capacity of the GA to propose a well-spaced front with minimum clustering.
The distance between each neighbor is somewhat constant, to the exception of the individuals located near the extremes.

Then, we perform a single-fidelity Kriging-based (SFK) optimization enriched with infills as described on the LF level in Section \ref{Infill strategy}.
We set the size of the DoE to $10\times N_p=40$, where $N_p=4$ is the number of design parameters, and the designs are chosen with a LHS.
After extensive testing, we find that an infill frequency of once every 20 generations over 100 generations is sufficient, amounting to 5 infill calls.
Then, we further converge the optimization with the SFK model only for another 50 generations (these parameters are summarized in Table \ref{tab: optim params}).
The resulting Pareto front has been added to Figure \ref{fig: Pareto RANS LF}, and corresponds well to the direct optimization outcome.
Finally, we recompute the SFK optimal designs with the LF RANS. 
The agreement is good as the surrogate model successfully predicts true Pareto optimal solutions, albeit the discontinuity in the true set is not resolved.

\subsection{MF co-Kriging (MFK) optimization}\label{MF co-Kriging optimization}
To train the MFK model, a DoE is built at both LF and HF levels.
On the LF level, the DoE described in the previous section is used, which we showed to be adequate for a SFK optimization.
On the other hand, a LHS with a very low number of HF LES observations (the exact number is discussed in the next paragraph) distributed among high- and low-performing designs may induce an unreasonably large surrogate error near the optimal region, altogether misguiding the optimization algorithm.
Therefore, we make the assumption that the RANS and LES samples are initially well-correlated, such that the least-performing candidates are reasonably well identified with the RANS approach on the LF level, allowing us to avoid computing such designs with LES.
Then, we build the HF DoE by extracting a set of Pareto-dominant  and extreme (low $J_1$ \& high $J_2$ and vice versa) designs from the LF DoE, plus the baseline configuration, which are recomputed using LES.

\begin{figure}[h!]
    \centering
    \includegraphics[width=0.49\textwidth]{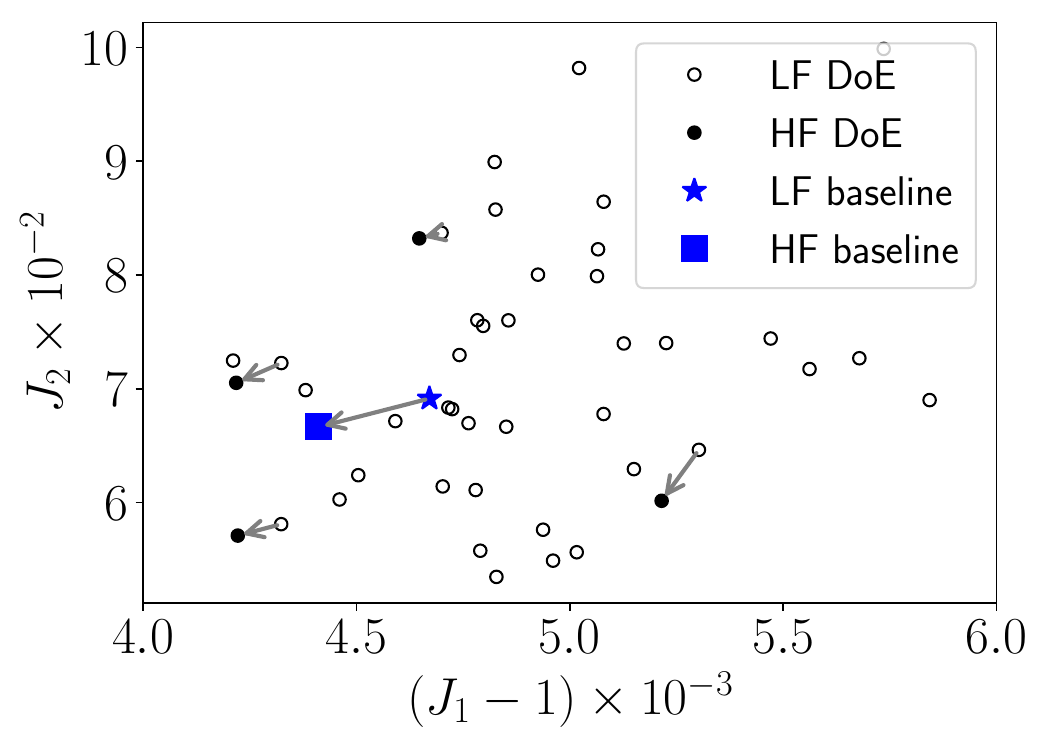}
    \includegraphics[width=0.49\textwidth]{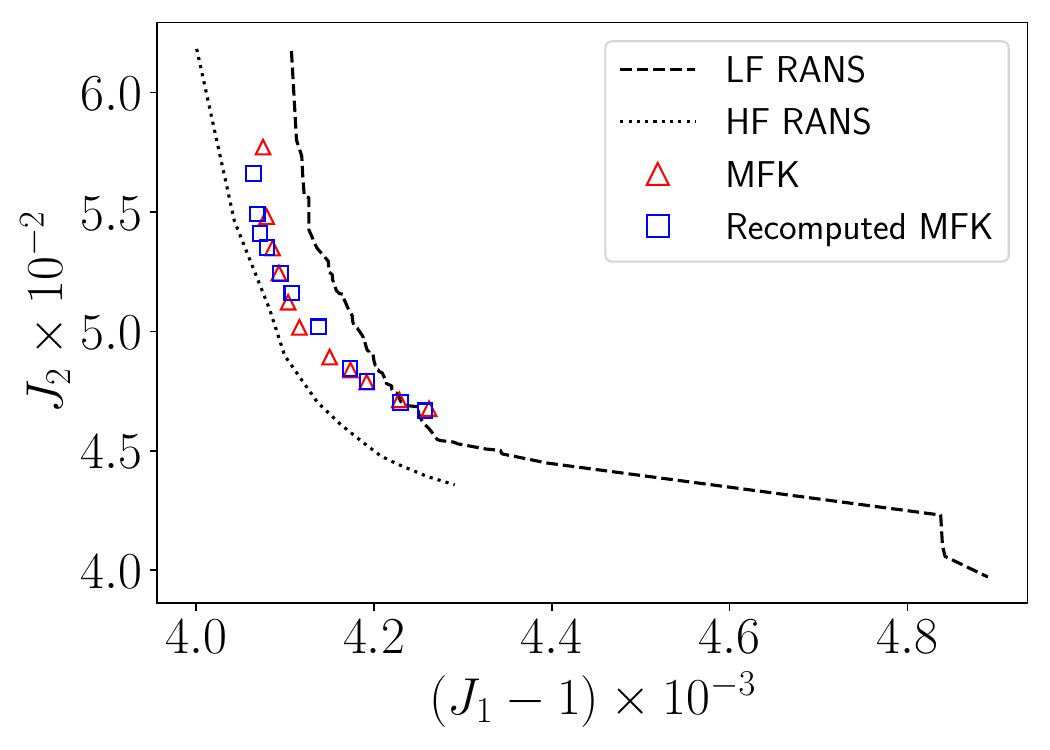}\\
    \makebox[0.49\textwidth][c]{a) LF and HF RANS DoE.}
    \makebox[0.49\textwidth][c]{b) MFK Pareto set.}
    \caption{a) LF and HF RANS DoE with arrows indicating the corresponding LF and HF samples, and b) Pareto set obtained with the LF, HF RANS and MFK optimizations, and recomputed with the HF RANS.}
    \label{fig: MF RANS DoE Pareto}
\end{figure}

In practice, a finite computational resource budget is available, and a compromise between the number of HF infills and DoE samples must be made. 
For this study, the total available budget was $1\,000\,000$CPUh. 
The cost of 1 LES with the present structured solver is $\approx130\,000$CPUh (440M points in the mesh, 4\,704 CPUs), thus only 8 LES can be run, of which 1 is used to validate 1 Pareto optimal design, and another is used to simulate the baseline configuration, leaving 6 LES to be distributed among the DoE and infills.
A series of tests was then performed using just RANS on two grid levels.
The coarse grid was obtained by retaining one every other point of the fine grid, leading to a LF and a HF RANS. The results indicate that a HF DoE consisting of 4 designs in addition to the baseline, and a total of 2 HF infills at half the frequency of the LF infill inform the MFK well enough to recover a solution close to the full HF RANS optimization (these parameters are summarized in Table \ref{tab: optim params}).
We show in Figure \ref{fig: MF RANS DoE Pareto}.a the chosen DoE samples of the HF RANS, where the arrows point from the corresponding LF observation.
Overall, the HF predictions appear somewhat scaled to lower $J_1$ and $J_2$ values.
Therefore, both fidelity levels are well correlated and marginal improvement is obtained over the LF function.
The resulting MFK Pareto front is given in Figure \ref{fig: MF RANS DoE Pareto}.b, where candidates have been recomputed using the HF RANS. 
Furthermore, designs from the LF RANS Pareto set were also recomputed with the HF RANS to approximate the true HF Pareto set.
The MFK predictions match with the designs true performance when recomputed with the HF function.
Furthermore, the true HF Pareto front is well approximated considering the number of total function evaluations with the HF RANS of only 7 and just 56 with the LF RANS.
Specifically, the MFK model correctly predicts the best compromise designs, although it slightly under-performs in the extrema regions.
Finally, the strategy described in Section \ref{Infill strategy} whereby the additional HF samples correspond to the PI computed with the LF, is also validated by the present tests.
\begin{table}
\begin{center}
    \caption{Summary of optimization parameters. $N_d$ is the number of designs per generation, $N_g$ the number of generations, $N_{l,i}$ and $N_{h,i}$ the numbers of LF and HF infills, $N_{l,total}$ and $N_{h,total}$ the number of total LF and HF methods calls.}
    \begin{tabular}{c|c|c|c|c|c|c|c}
        \textbf{Run} & $N_d$ & $N_g$ & $N_{l,i}$ ($\times3$) & $N_{h,i}$ & Last infill generation & $N_{l,total}$ & $N_{h,total}$ \\ \hline
        LF RANS & 30 & 70 & - & - & - & 2\,100 & - \\ 
        SFK & 30 & 150 & 5 & - & 100 & 56 & - \\ 
        MFK & 30 & 150 & 5 & 2 & 100 & 56 & 7 \\ 
    \end{tabular}
    \label{tab: optim params}
\end{center}
\end{table}%


\section{Results of the multi-fidelity LES-RANS optimization}
\subsection{Pareto front analysis}\label{Functional space}
\begin{figure}[h!]
	\centering
	\includegraphics[width=0.5\textwidth]{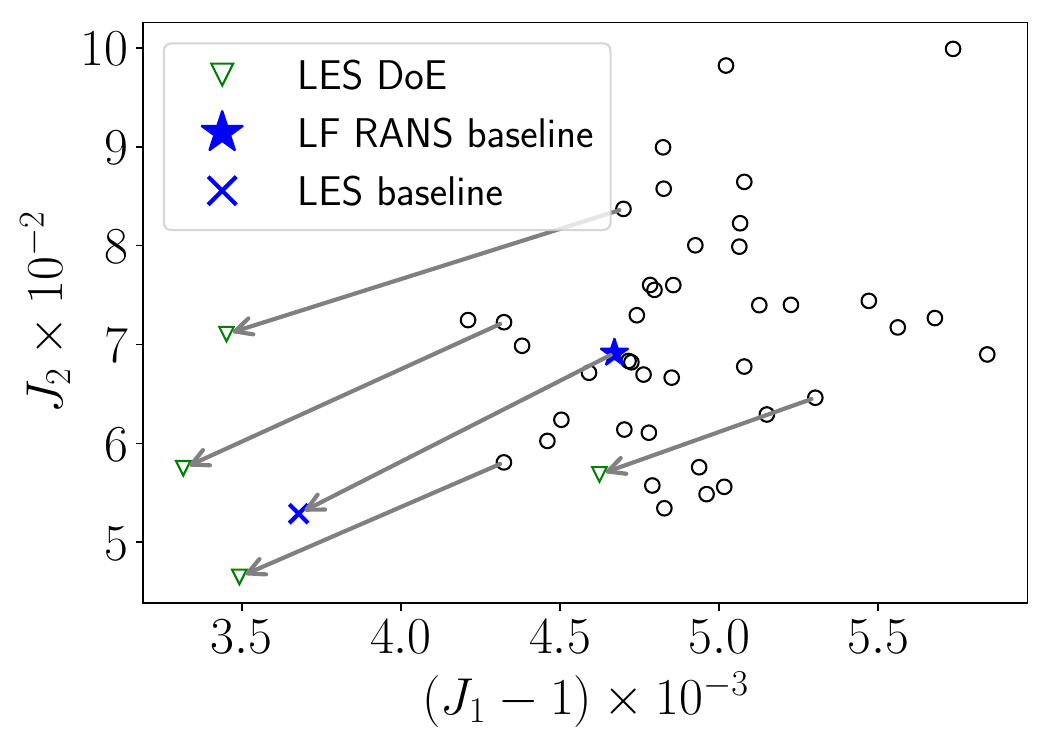}
    \caption{RANS and LES DoEs with arrows indicating the corresponding LF and HF samples.}
    \label{DoE LES}
\end{figure}%
\begin{figure}[h!]
	\centering
	\includegraphics[width=0.99\textwidth]{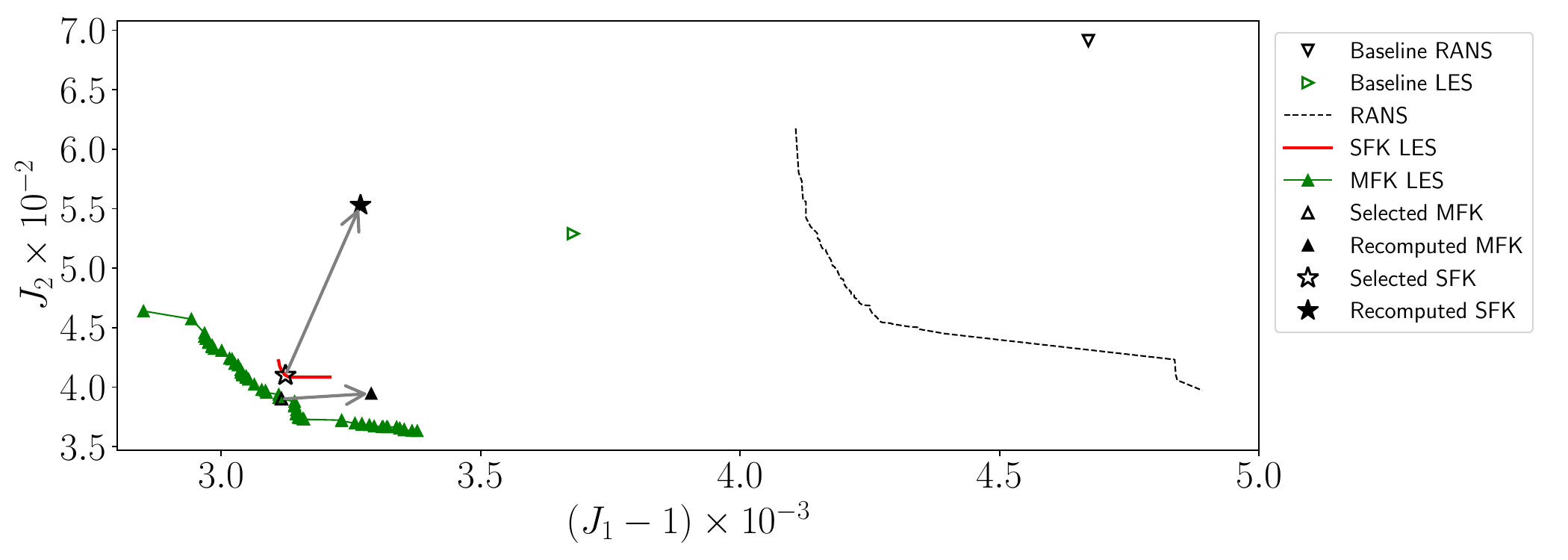}
    \caption{Comparison of predicted Pareto fronts from several optimizations.}
    \label{LES results}
\end{figure}%
We now discuss the multi-fidelity optimization using the LES and RANS methods.
We show the performance of the LES DoE designs in Figure \ref{DoE LES}. 
The arrows point from the corresponding RANS predictions. 
Evidently, the LF computations largely overestimate $J_1$ and to a lesser extent $J_2$, which was anticipated in the observations made in Section \ref{RANS and LES predictions}. 
Nonetheless, we observe a reasonable correlation between both levels of fidelity which initially supports the MFK model founding hypothesis, as well as the present assumption of constant adjustment coefficient $\rho_{l}(\mathbf{d})=\rho_{l}$.
We show in Figure \ref{LES results} the resulting LES Pareto front predicted with the MFK model.
First, we observe that the MFK model does provide a family of better performing designs compared to the baseline evaluated with the LES.
We then select a Pareto optimal design (Selected MFK) and recompute it with the last LES available (Recomputed MFK), and its true performance are given in the same figure.
Good agreement between the surrogate prediction and the true performance is achieved considering the overall low number of HF and LF calls, resulting in deviations of $\approx5.6\%$ and $\approx2.4\%$ in $J_1-1$ and $J_2$ objective functions, respectively.
Comparatively, the RANS Pareto designs have an overall minimum error with respect to the selected LES of $\approx24.6\%$ and $\approx12.7\%$ in $J_1-1$ and $J_2$, respectively, translating into a reduction in error by factors of about $\approx4.4$ and $\approx5.3$.

Interestingly, the MFK surrogate based on the two fidelity levels of RANS computations discussed in Section \ref{MF co-Kriging optimization} appeared more proficient at predicting optimal designs of the HF RANS, when compared to the surrogate used for the LES.
Since both optimizations rely on the same DoE size and quantity of infills, this hints that the number of HF samples (be they RANS or LES) is not the sole driving factor behind the performance of the MF surrogate.
Rather, the correlation between the LF and HF samples is the most dominant parameter, which in the MFK model is encapsulated within the adjustment coefficient $\rho_{l}(\mathbf{d})$.
In this view, increasing the DoE size allows to compensate for a poor correlation between two levels of fidelity, while fewer samples are required to efficiently train a MF surrogate on highly correlated data.
On the one hand, the latter case was verified in the MF RANS optimization as the LF and HF RANS predictions are very well correlated, on account of the fidelity being quantified by the level of grid refinement which only seemed to slightly alter the solution.
There, assuming $\rho_{l}(\mathbf{d})=\rho_{l}$ is relevant as only a simple scaling is required to correct most of the LF RANS predictions.
On the other hand, the \textit{a priori} strong correlation between the RANS and LES DoE observed in Figure \ref{DoE LES}.b may have in fact deteriorated as the optimization procedure reached the optimal region.
This suggests that the present RANS methodology is capable of capturing the global trends of blade performance variation with large shape deformations, while it fails at correctly predicting the impact of fine geometric perturbations on blade losses.
Furthermore, while the test series on the LF and HF RANS allowed to estimate the optimization and Kriging model parameters, they are not truly representative of the practical issues encountered when dealing with both low-cost steady computations and expensive scale-resolving simulations.

In the MFK model, the LF RANS serves to explore the design space at a fraction of the cost of the LES, and its predictions are subsequently corrected by the model.
However, if the correlation between the RANS and LES is weak and the amount of data to allow for more complex forms of the adjustment coefficient $\rho_{l}(\mathbf{d})$ (linear or quadratic, rather than constant) is insufficient, one could suspect the MFK approach to under perform when compared to a SF model solely based on the very few LES samples.
Therefore, we verify this claim by building a SFK model trained on the same LES DoE, augmented by the two LES infills computed during the MFK optimization.
The resulting Pareto front (SFK LES) is included in Figure \ref{LES results}.
Although it lies close to the MFK LES Pareto front, it only covers a very small region of the functional space.
We then select a Pareto optimal design (Selected SFK) and recompute it with LES (Recomputed SFK).
Note that this additional computation was not accounted for within the budget declared in Section \ref{MF co-Kriging optimization} and ensued an overhead of $\approx130\,000$CPUh.
The SFK prediction error is large as the true performance are offset by $\approx4.9\%$ and $\approx25.5\%$ in $J_1-1$ and $J_2$, respectively.
We thus conclude that the RANS and LES predictions are sufficiently correlated for the MFK model to efficiently model their correlation under the assumption of constant $\rho_{l}$.\\

\begin{figure}[h!]
    \centering
    \includegraphics[width=0.99\textwidth]{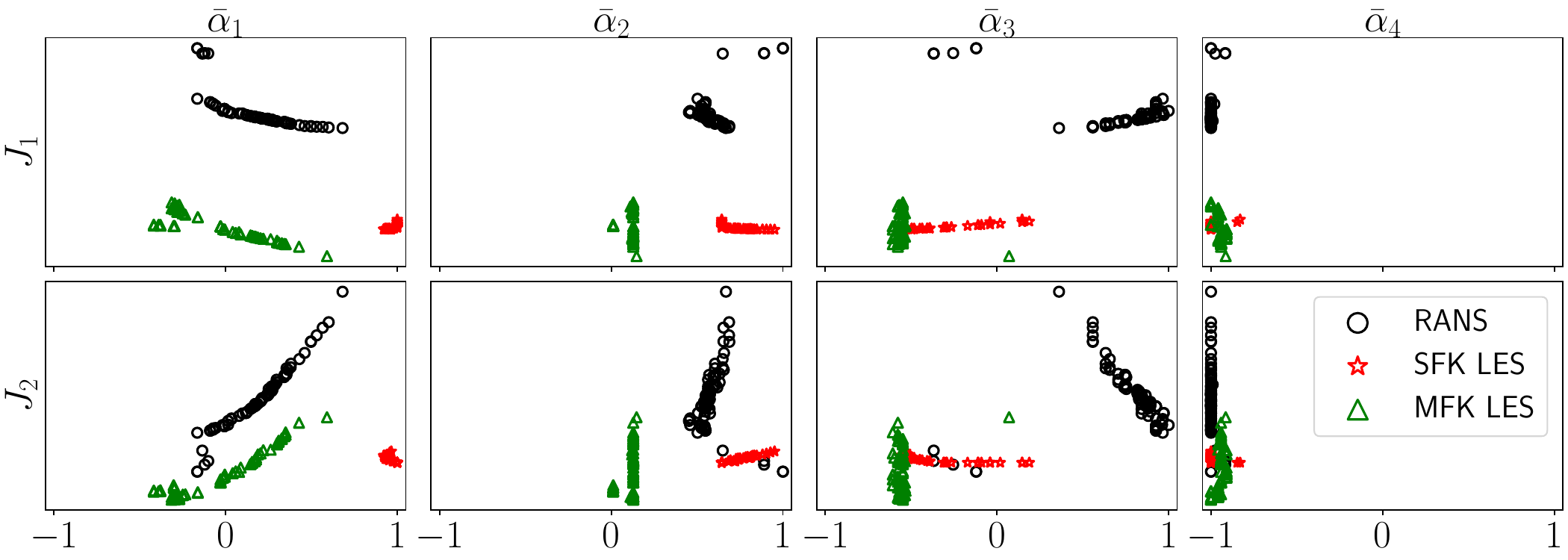}
    \caption{Comparison of predicted Pareto fronts in the design parameter space.}
    \label{fig: Pareto_parameter_space}
\end{figure}%

We now inspect the various predicted Pareto sets in the parameter space.
We plot in Figure \ref{fig: Pareto_parameter_space} the objective functions against each normalized design parameter.
The latter are the normalized SVD modal coefficients $\Bar{\alpha}_i$ which control the contribution of each mode (shown in Figure \ref{fig: modes}) to the shape perturbations.
We first compare the RANS and MFK LES optimal design spaces.
Both methods identify very similar optimal first modal coefficients $\Bar{\alpha}_1$, of which the associated mode essentially controls the blade camber.
Especially, the trends are alike and positively correlated.
According to the RANS however, a larger portion of the optimal $\Bar{\alpha}_1$ is positive, favoring blades with positive camber.
On the other hand, the LES predictions are more nuanced and provide additional negatively cambered blades.
Then, the initially strong correlation between the RANS and LES completely vanishes for the next two optimal modal coefficients.
Specifically, a somewhat fixed value of $\Bar{\alpha}_{2-3}$ is obtained with the LES, which is not corroborated by the RANS.
Furthermore, the RANS Pareto optimal $\Bar{\alpha}_{2-3}$ are again positive which further contributes to deform and raise the LE, as hinted by the mode shapes in Figure \ref{fig: modes}.
We further discuss this particular point in Section \ref{Role of LE and TE in the LES}, as we will show that it plays a crucial role in the present blade performance.
To conclude, both approaches identify very similar optimal global geometric trends such as blade camber, but diverge on the role of finer geometric details.

We also include the results from the SFK LES optimization, and observe that it identifies an optimal region without apparent correlation to either the RANS of MFK LES trends.
This further supports the claim that the MFK LES has indeed benefited from the RANS computations, which correctly predict the trends at least with respect to the first SVD mode, although we suspect that the correlation between the two levels of fidelity may have deteriorated as the optimal region was approached.

\subsection{RANS and LES designs comparison}\label{RANS and LES designs comparison}
\begin{figure}[h!]
    \centering
    \includegraphics[width=0.49\textwidth] {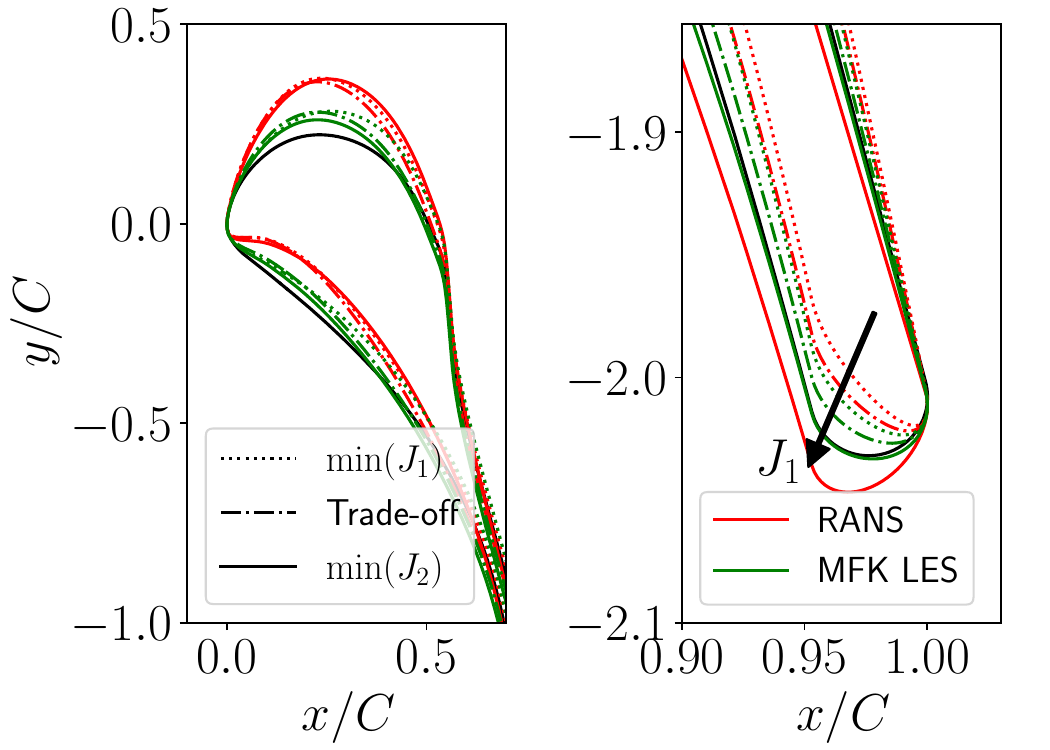}
    \includegraphics[width=0.49\textwidth]{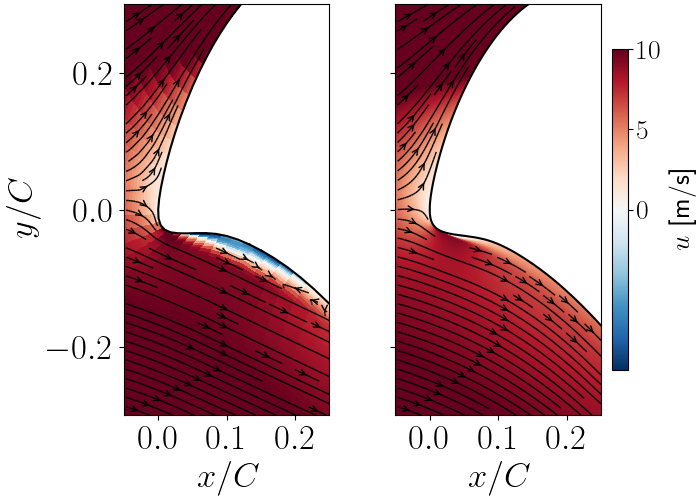}\\
    \caption{a) Pareto optimal geometries predicted by both RANS and LES MF co-Kriging surrogates, b) streamwise velocity field inside (left) RANS Pareto optimal and (right) LES worst DoE geometries.}
    \label{Comp geoms}
\end{figure}%
We now compare 3 Pareto optimal designs predicted by the RANS and MFK LES optimizations in Figure \ref{Comp geoms}.a, consisting in the minimum of each objective function and the most centered along the front.
Interestingly, the correlation of the increasing deformation of the LE and thinning TE with the decrease (increase) in $J_1$ ($J_2$) is similar between both levels of fidelity. 
Furthermore, this trend indicates that the Mach number field downstream of the blade becomes more uniform with thicker TE, whereas the associated entropy generation is enhanced. 
We note however that the overall deformations are milder in the LES optimal designs, as anticipated by the overall lower modal coefficients shown in Figure \ref{fig: Pareto_parameter_space}.

We report in Figure \ref{Comp geoms}.b the streamwise velocity field in the LE vicinity inside a RANS Pareto optimal (left) design, that we compare to another very similar blade (right) for which an LES was performed during the MFK optimization, and that is sub-optimal (part of the DoE).
We note that both geometries are deformed in a similar fashion and feature a strong curvature on the pressure side.
This causes boundary layer separation leading to a large recirculation bubble in the RANS, and to a lesser extent in the LES.
Surprisingly, this is a common trait shared by all RANS Pareto optimal candidates, which indicates that the additional entropy generated by the flow separation and recirculation only plays a minor role in the performance reduction of the blade when computed with the present RANS.
This is made possible by the underlying assumption of fully turbulent boundary layer made in the present Spalart-Allmaras turbulence model formulation, which causes global losses largely surpassing that of the LE separation.
On the other hand, the much smaller separated flow region in the LES sub-optimal design is sufficient to reduce the performance of the blade compared to other candidates computed with LES.
This point is elaborated on in the next section.
\begin{figure}[h!]
    \centering
    \includegraphics[width=0.49\textwidth]{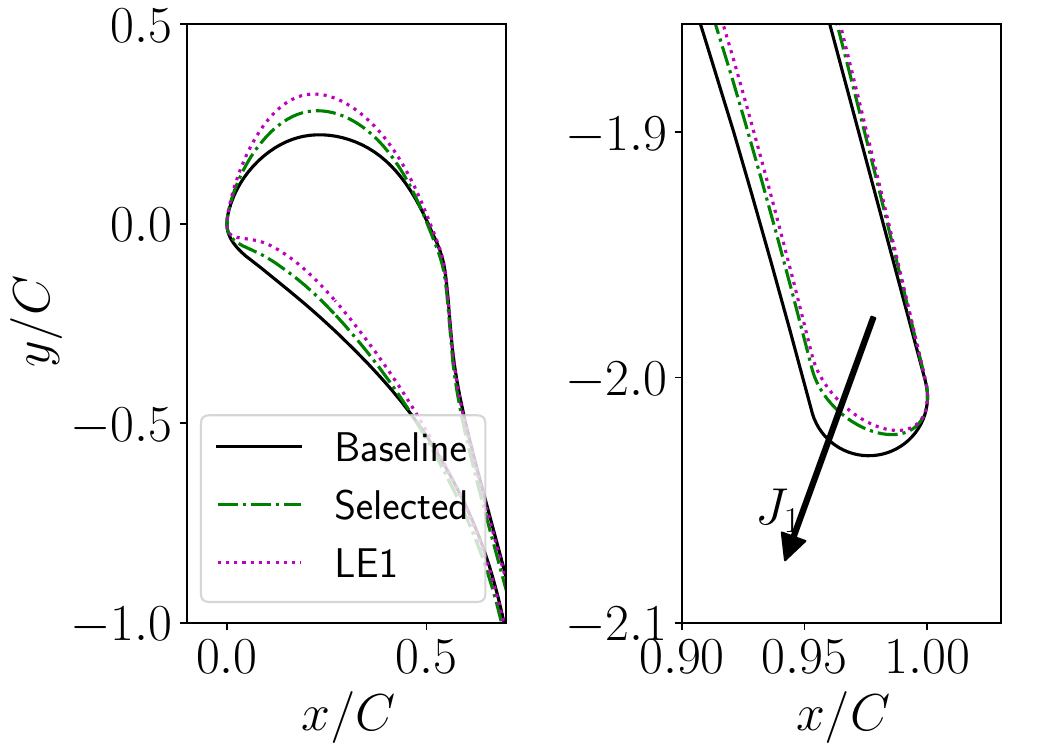}
    \includegraphics[width=0.49\textwidth]{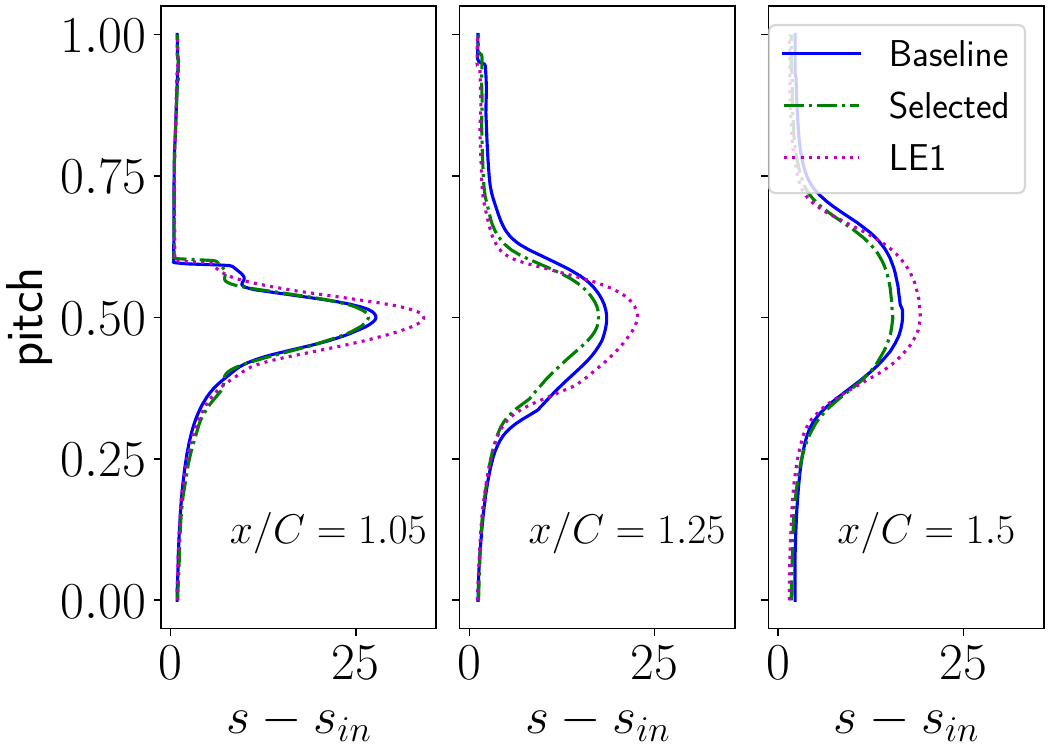}\\
    \caption{a) LES designs details around the LE and TE, b) entropy deviation profiles across the LES wakes at three locations downstream of the blades.}
    \label{Comp LES}
\end{figure}%

\subsection{Role of LE and TE in the LES}\label{Role of LE and TE in the LES}
To illustrate the role of the LE and TE topology on blade performance as predicted by the LES, we compare three geometries: the Selected (which served to validate the MFK model, see Figure \ref{LES results}), the sub-optimal design (discussed in the previous paragraph) from hereon denoted as LE1, and the Baseline.
We note that these candidates dominate each other in that respective order, and provide their objective function values in Table \ref{tab: LES validation}.
We show their geometries in Figure \ref{Comp LES}.a.
In the following discussion, we expose the physical processes behind why the highly deformed LE topology (such as the LE1 blade) was not retained during the LES optimization cycle, while it populates completely the RANS Pareto front.

\begin{table}
\caption[Table]{Performance of LES designs chosen for in-depth performance analysis}\label{tab: LES validation}
\centering{%
	\begin{tabular}{l|c|c}
		
		\textbf{Design} & $(J_1-1)\times10^{-3}$ & $J_2\times10^{-2}$ \\
		\hline
		Baseline & 3.679 & 5.290 \\
		LE1 & 3.492 & 4.651 \\
		Selected & 3.289 & 3.945 \\
	\end{tabular}
}
\end{table}%

\begin{figure}[h!]
	\centering
	\includegraphics[width=1\textwidth]{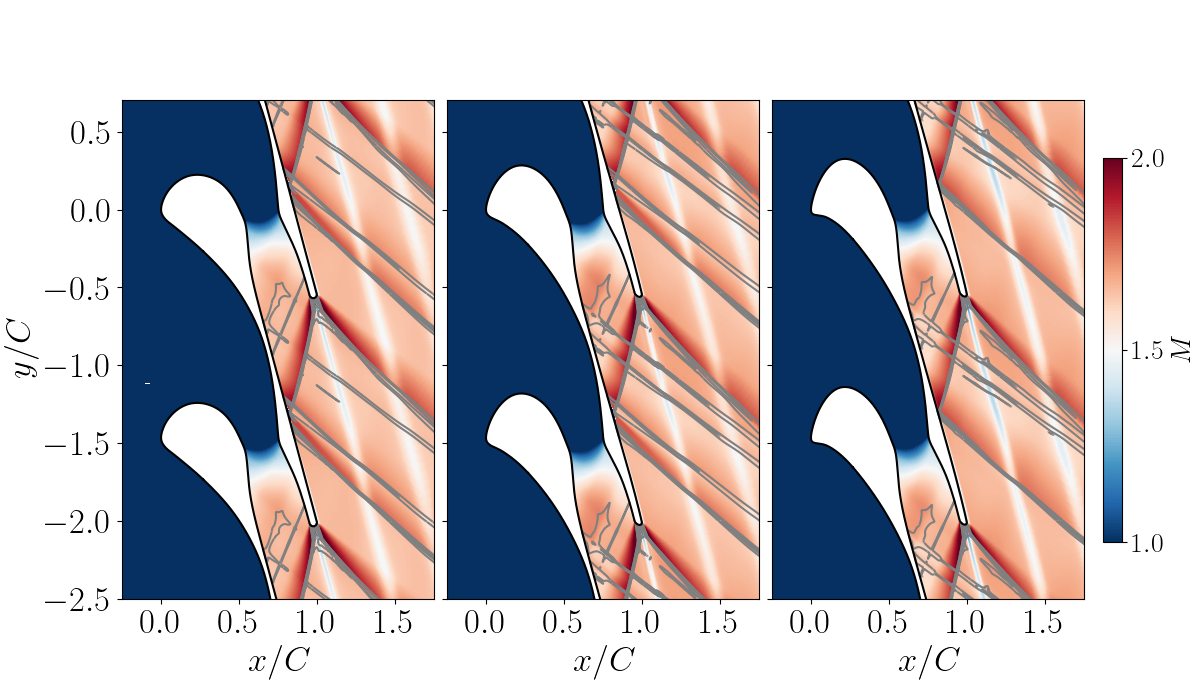}
	\makebox[0.30\textwidth][c]{a) Baseline.}
	\makebox[0.30\textwidth][c]{b) Selected.}
	\makebox[0.30\textwidth][c]{c) LE1.}
	\caption{LES optimization chosen candidates mean Mach number fields and velocity divergence contours.}
	\label{fig: comp LES M}
\end{figure}

We first provide a qualitative description of the flow and show the mean Mach number fields and velocity divergence contours from each computation in Figure \ref{fig: comp LES M}.
The contours allow to clearly outline the shock and Mach wave patterns.
In addition to the shock wave system located at the TE, clustered contours are clearly visible upstream inside the passage for all designs.
In the Baseline geometry, this stems from Mach waves generated on the pressure side (PS) (the angle computed for an ideal gas flow with local Mach number of 1.5 is $\approx42$\textdegree and agrees with this observation) due to the growing boundary layer, which reflect on the suction side (SS).
Note that the boundary layer on the Baseline PS is fully laminar until the TE (this is shown in the next paragraphs).
Then, at the same location, both better performing candidates (Selected and LE1) feature more pronounced compression waves which highlight a distinct reflection pattern on the blade SS.
This may stem from the nozzle divergent section deformations, strengthening the initial small perturbations of the Mach waves.
Furthermore, one notable difference between the Selected and LE1 designs lies in the TE shock/SS boundary layer interaction, which only features one reflected shock in the LE1 case.
In SWBLIs, the shock pattern is mainly controlled by the state of the incoming boundary layer, whereby laminar boundary layers are not able to resist the sharp pressure gradient, leading to the formation of a large recirculation bubble, an expansion fan and additional shock waves \cite{Sansica2015}.
On the other hand, the typical shock system around turbulent boundary layers consists in a much smaller or inexistent recirculation bubble, leading to the coalescence of the shock system to just one visible reflected shock.
Therefore, the SS boundary layer in the LE1 geometry may either be transitional or fully turbulent upstream of the SWBLI, possibly as a result of the aforementioned upstream compression waves, of even greater strength compared to the Selected geometry.
Finally, the wakes appear mildly affected, although a more pronounced velocity deficit is observed in the LE1 case.

We now compare the wake entropy profiles at several locations downstream of the blades in Figure \ref{Comp LES}.b.
The widest wake is obtained in the Baseline design which is attributable to its TE being the thickest.
Then, the strongest entropy content is reached in the LE1 design, which also produces the thinnest wake.
Interestingly, the Selected design produces a slightly thicker wake compared to the LE1 and yet generates the less entropy overall.
Therefore, the maximum entropy content inside the wake does not vary monotonically with TE thickness.
In fact, we will show that the deformations of the nozzle divergent section and of the LE play a major role in the entropy production inside the wake.

\begin{figure}[h!]
    \centering
    \includegraphics[width=0.49\textwidth]{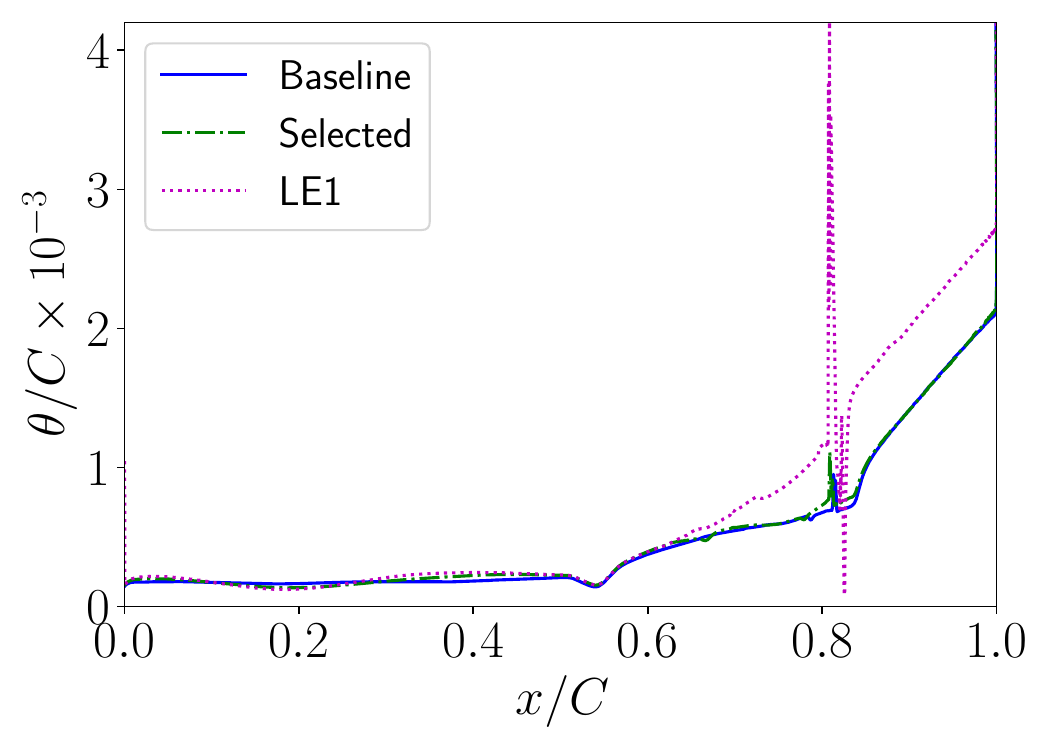}
    \hfill
    \includegraphics[width=0.49\textwidth]{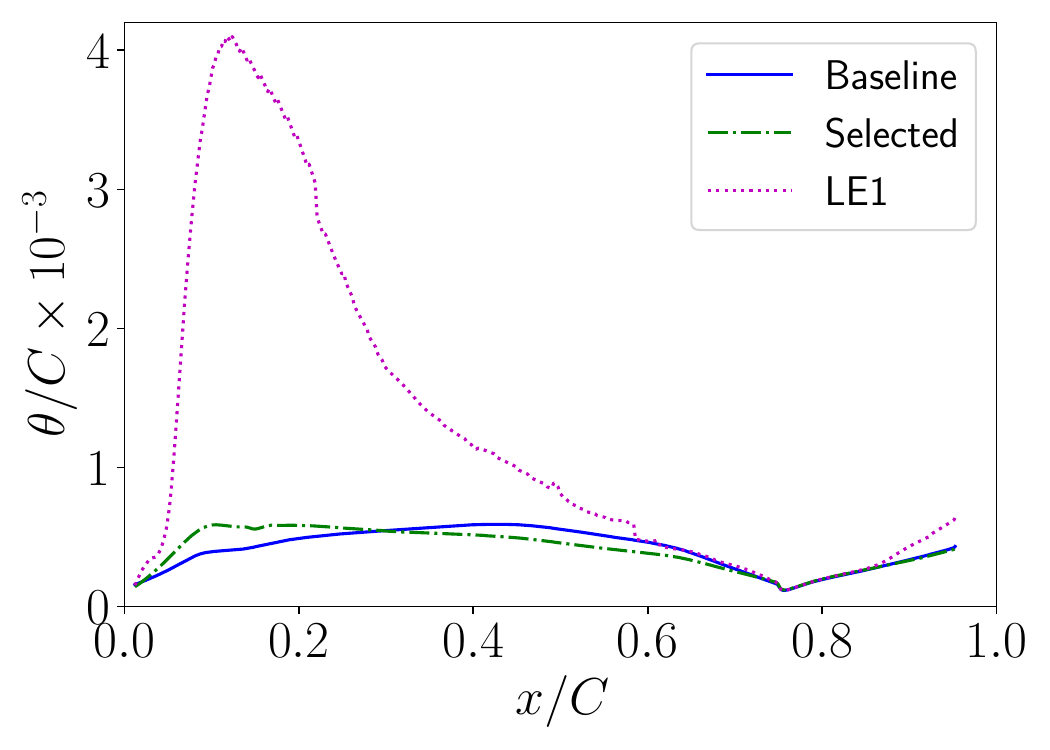}\\
    \makebox[0.49\textwidth][c]{a) Suction side.}
    \makebox[0.49\textwidth][c]{b) Pressure side.}
    \caption{a) Boundary layer displacement thickness evolution against axial chord from the Baseline, Selected and LE1 geometries.}
    \label{fig: theta comp LES}
\end{figure}

\begin{figure}[h!]
    \centering
    \includegraphics[width=0.49\textwidth]{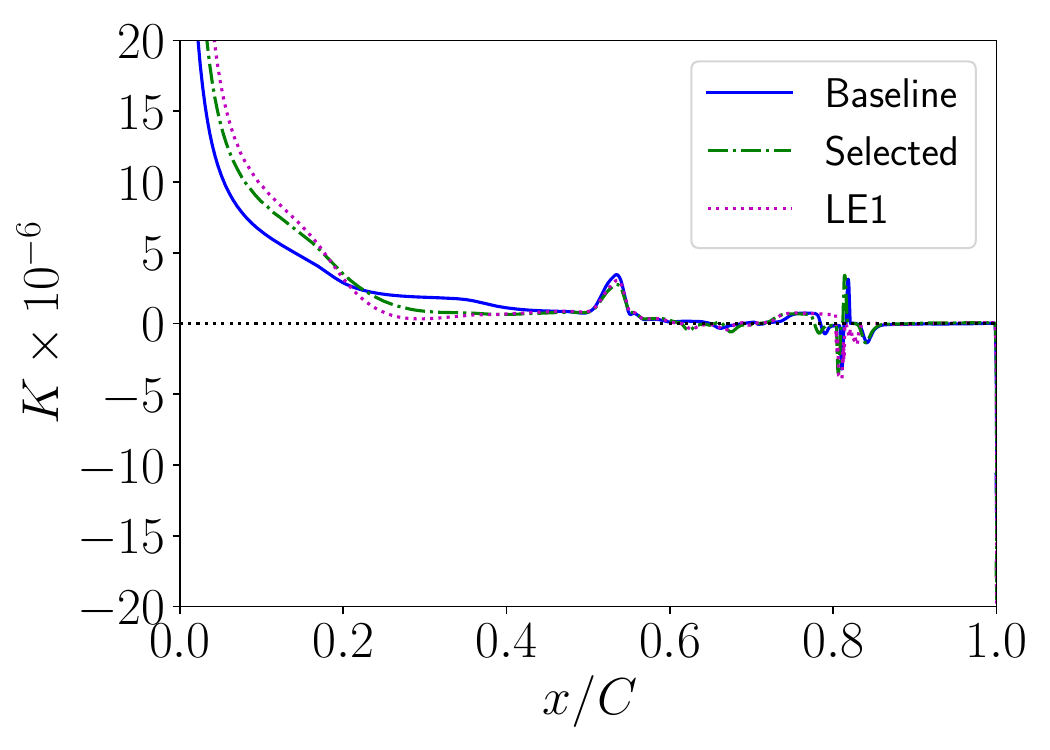}
    \hfill
    \includegraphics[width=0.49\textwidth]{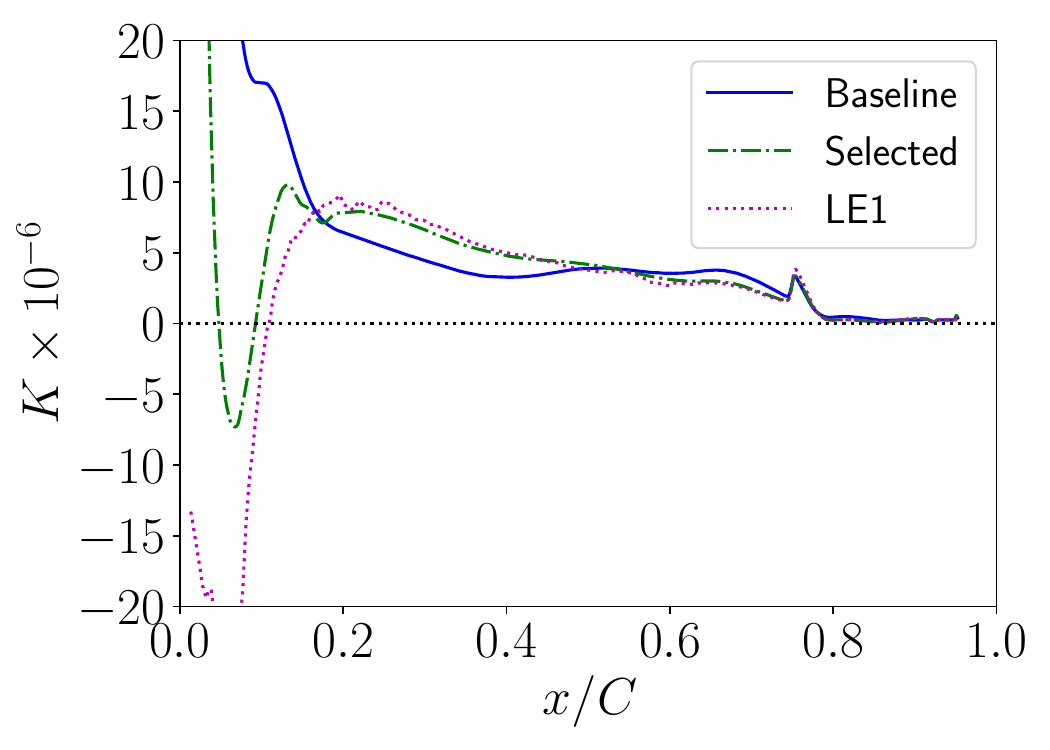}\\
    \makebox[0.49\textwidth][c]{a) Suction side.}
    \makebox[0.49\textwidth][c]{b) Pressure side.}
    \caption{a) Acceleration parameter evolution against axial chord from the Baseline, Selected and LE1 geometries.}
    \label{fig: K LES}
\end{figure}

In the RANS computations, the highly deformed LE causing boundary layer separation on the PS had no impact on the Pareto optimal designs.
In the LES however, the separation strongly contributes to reduce the vane performance.
To illustrate this, we first report in Figure \ref{fig: theta comp LES} the boundary layer momentum thickness $\theta$ evolution against axial location on both SS and PS.
To begin with, we describe the Baseline configuration.
On the SS, $\theta$ remains approximately constant until the end of the divergent section of the nozzle, located at $x/C\approx0.6$.
This is typical of accelerated boundary layers, and we provide the acceleration parameter $K=\frac{\nu}{U_e}\frac{\mathrm{d}U_e}{\mathrm{d}x}$ in Figure \ref{fig: K LES}.a to support this claim.
Correspondingly, $K$ remains positive (favorable pressure gradient) until $x/C\approx0.6$ which coincides with the end of the divergent and the flow expansion.
Finally, the increase in momentum thickness after $x/C\approx0.8$ is caused by the SWBLI triggering boundary layer transition. 
On the PS, $\theta$ increases to a maximum before reducing to a minimum at $x/C\approx0.75$.
Similarly, this behavior can be linked to $K$ in Figure \ref{fig: K LES}.b which also remains positive.
In the case of the Selected geometry, the overall evolution of $\theta$ is comparable to that in the Baseline vane, only $K$ on the PS is negative on the LE which initially enhances the momentum thickness.

\begin{figure}[h!]
    \centering
    \includegraphics[width=0.49\textwidth]{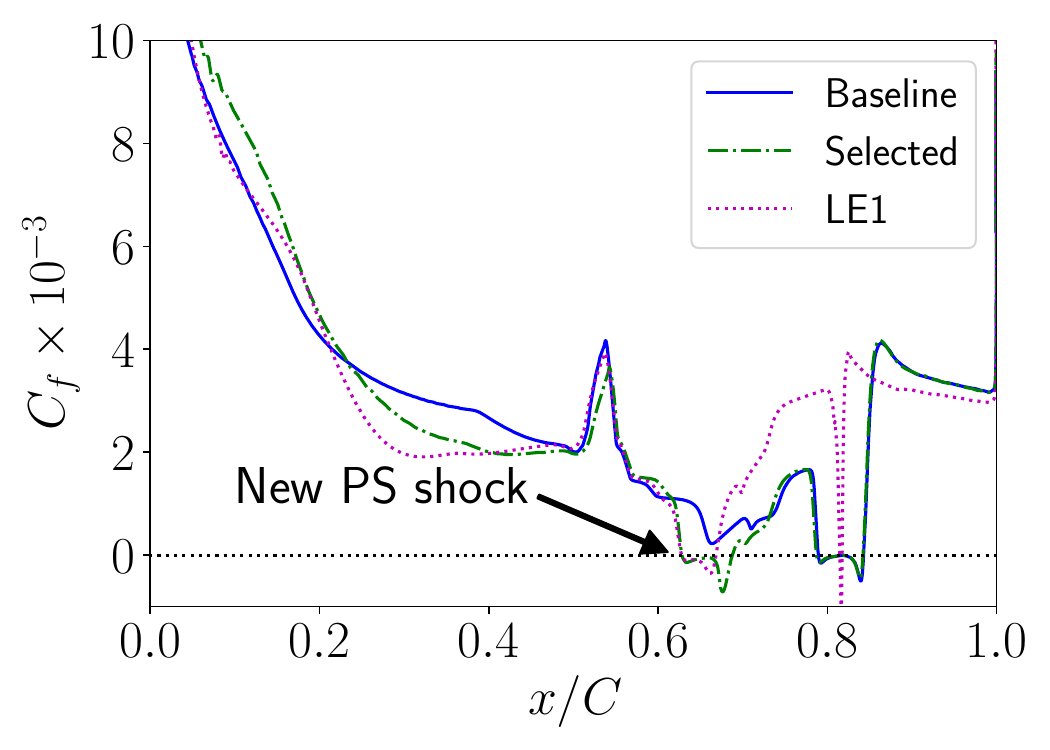}
    \hfill
    \includegraphics[width=0.49\textwidth]{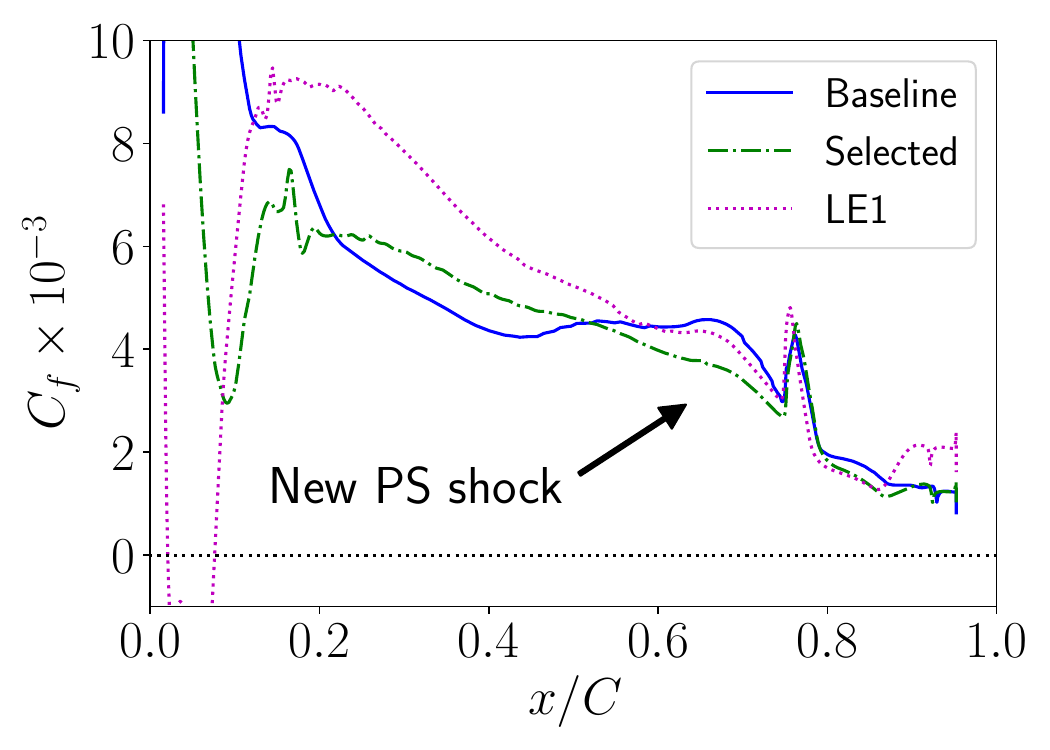}\\
    \makebox[0.49\textwidth][c]{a) Suction side.}
    \makebox[0.49\textwidth][c]{b) Pressure side.}
    \caption{a) Friction coefficient evolution against axial chord from the Baseline, Selected and LE1 geometries.}
    \label{fig: Cf comp LES}
\end{figure}

In contrast, important deviations are obtained in the LE1 design on both SS and PS.
On the former, a departure of $\theta$ from the other two cases begins slightly downstream of the divergent.
This location corresponds specifically to the impingement of the strengthened PS Mach waves compared to the Baseline (discussed at the beginning of the present section), which we will denote as "new PS shock" for conciseness.
Indeed, the friction coefficient $C_f$ (shown in Figure \ref{fig: Cf comp LES}) assumes a negative value over $\approx5\%C$ in the region $x/C\approx0.65$ on the SS as a result of the interaction between the boundary layer and the new PS shock (indicated in the Figure).
This in turn indicates separation leading to a recirculation bubble, implying that the incoming boundary layer is laminar.
Subsequently, transition to turbulence is triggered and both $\theta$ and $C_f$ rise accordingly.
Similar comments can be made for the Selected design, although the negative $C_f$ region found at the same location extends further downstream, showing that the boundary layer reattaches at a later stage, after which it remains laminar as in the Baseline design.
This indicates that, in the LE1 design, the boundary layer transitions to turbulence inside the recirculation region, causing the earlier reattachment and subsequent rise in $C_f$ and $\theta$.
These two dissimilar behaviors may ultimately be traced back to the strength of the new PS shock, which we suspect is milder in the Selected design compared to the LE1.
Finally, the largest discrepancies in momentum thickness clearly arise on the PS, due to the boundary layer separation just aft the LE which affects the near-wall flow over $60\%$ of the axial chord.
The corresponding $K$ (Figure \ref{fig: K LES}.b) presents a large negative region, which hints that the local adverse pressure gradient due to the strong curvature of the geometry is responsible for the boundary layer separation.

\subsubsection{Entropy generation inside the boundary layers}\label{Entropy generation inside the boundary layers}
\begin{figure}[h!]
    \centering
    \includegraphics[width=0.49\textwidth]{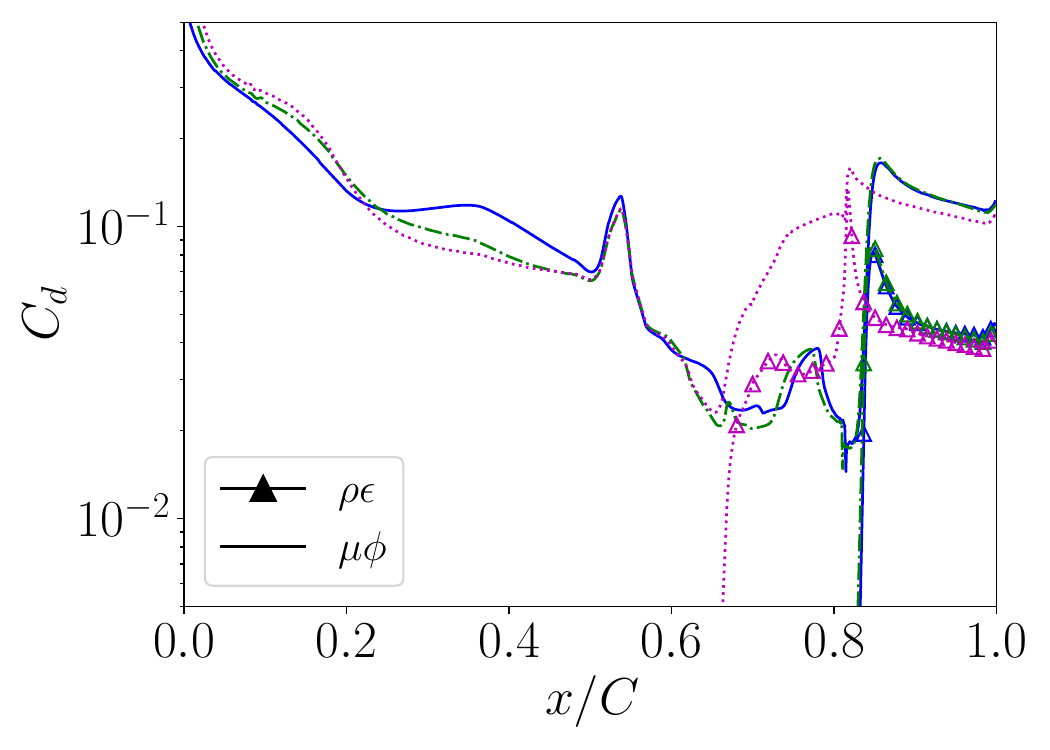}
    \hfill
    \includegraphics[width=0.49\textwidth]{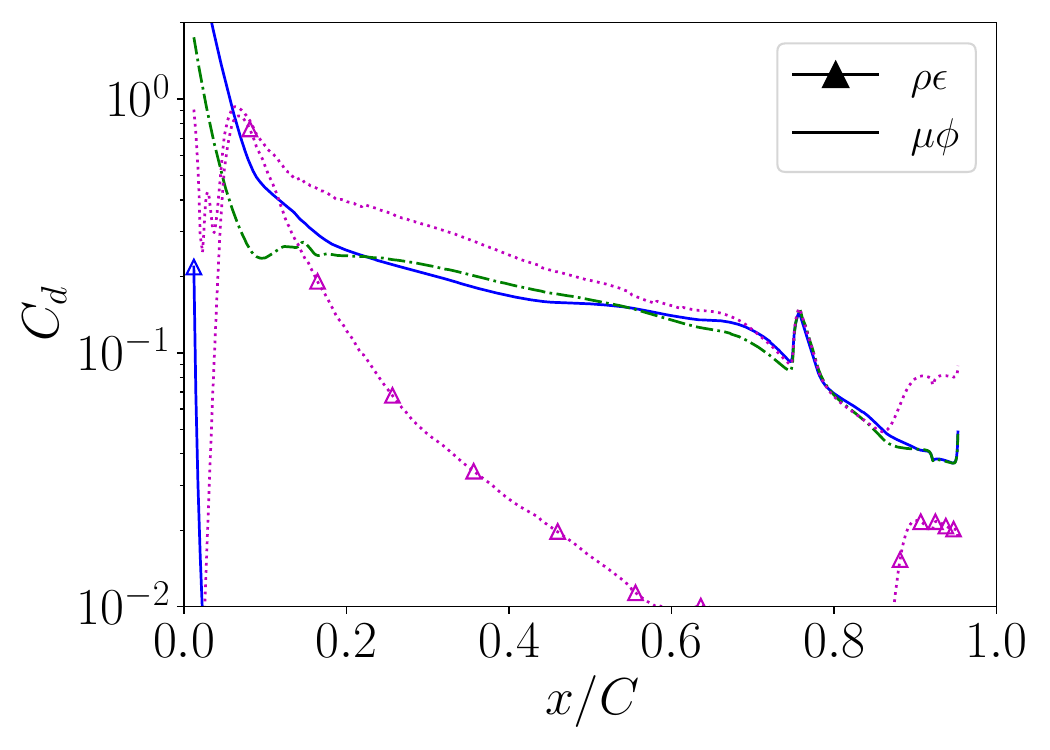}\\
    \makebox[0.49\textwidth][c]{a) Suction side.}
    \makebox[0.49\textwidth][c]{b) Pressure side.}
    \caption{a) Boundary layer dissipation coefficient evolution against axial chord from the Baseline, Selected and LE1 geometries.}
    \label{fig: Cd split LES}
\end{figure}%
Now that we have drawn a clear picture of the various complex flow phenomena at play inside the chosen vane geometries, we inspect the resulting mean production rate of entropy inside the boundary layers.
To this end, we compute the dissipation coefficient $C_d$ \cite{Denton1993} on the three blades, defined as
\begin{equation}
    C_d = \frac{T_e\dot{S}''}{\rho_e U_e^3} = \mu\Phi + \rho\epsilon 
\end{equation}%
where the $C_d$ has been split into its 2 contributions $\mu\Phi$ and $\rho\epsilon$ as a result of mean and unsteady dissipation, respectively (see \cite{Walsh2011} for details).
$T_e$, $\rho_e$ and $U_e$ are the boundary layer edge temperature, density and velocity, and the integrated pointwise entropy production rate is given by
\begin{equation}\label{Sppp}
    \dot{S}''=\int_{0}^{\delta}\dot{S}'''\mathrm{d}y \quad\quad\quad\text{where}\quad\quad\quad \dot{S}''' = \overline{\tau_{ij}\frac{\partial u_i}{\partial x_j}}
\end{equation}%
where $\delta$ is the boundary layer height, $\tau_{ij}$ is the viscous stress tensor and $\Bar{\cdot}$ denotes a time- and span-averaged quantity.
By splitting $C_d$ into its mean and unsteady contributions, regions where no apparent $\rho\epsilon$ contribution is observed are deemed laminar.
We note that local unsteady content does not directly translate into turbulence, for instance the laminar wake behind a cylinder at low Reynolds number.
However, a direct link between $\rho\epsilon$ and turbulence can be made in attached flow regions \cite{Ghasemi2014} as the latter would be the only source of unsteadiness.
We show the evolution of the $C_d$ components in Figure \ref{fig: Cd split LES}.
The boundary layers inside all three vanes remain laminar over a large portion of the blade SS (Figure \ref{fig: Cd split LES}.a) until $x/C\approx0.65$, although the $C_d$ is slightly reduced in some instances in the optimized geometries (for instance in $x/C\in[0.2;0.5]$).
A strong rise in $\mu\Phi$ is then recorded for the LE1 case only, accompanied by the onset of unsteady dissipation $\rho\epsilon$.
This clearly supports the previous observations made with the $C_f$ that the boundary layer is no longer fully laminar.
Most importantly, entropy generation is enhanced over the SS and contributes to the deterioration of blade performance compared to the Baseline and Selected designs.
Then, the SWBLI resulting from the TE shock at $x/C\approx0.8$ leads to a fully turbulent state in all cases, and both components of $C_d$ have largely increased.

Likewise, the PS of the LE1 design features an overall higher level of $\mu\Phi$ compared to the other two, as well as a strong $\rho\epsilon$ contribution near the LE due to the flow separation.
Furthermore, the sharp rise in both contributions just upstream of the TE near $x/C=0.9$ caused by the LE separation evidences the strong history effects inside the boundary layer.
This is also in line with the previous observations of increasing $\theta$ and $C_f$ at that location (see Figures \ref{fig: theta comp LES}.b and \ref{fig: Cf comp LES}.b).
Overall, the resulting additional entropy generated on both SS and PS of the LE1 design is the footprint of complex mechanisms that have a direct impact on the loss level measured in the wake (see Figure \ref{Comp LES}.b).
While this renders the LE1 design sub-optimal, the latter remains an improvement over the Baseline simply due to its much thinner TE.

As expected, these complex flow phenomena are not resolved nor modeled by the present RANS strategy.
Therefore, the strong modeling assumptions and misrepresentations of loss generating mechanisms guide the optimization algorithm towards erroneous design solutions.

\section{Conclusion}
In this work, we performed shape optimization of a high-pressure turbine vane using a multi-fidelity co-Kriging (MFK) surrogate of both steady RANS computations and wall-resolved LES. 
The method for generating an appropriate Design of Experiments (DoE) for both levels was described, along with the infill strategy.
The single- (SFK) and MFK approach were validated, the latter using RANS only where the high-fidelity (HF) objective function evaluation consisted in a fine mesh computation, and the low-fidelity (LF) in a coarse mesh approximation.
This allowed to estimate optimal co-Kriging parameters, such as the frequency of LF and HF infills and the number of generations required to reach convergence.
A total of 7 LES were performed to train the MFK model, while 56 LF RANS samples were generated.
The surrogate successfully captured the correlation between the LES and RANS observations, and provided a good approximation of the better performing designs.
Furthermore, it showcased the disparities in optimal parameter spaces between LES and RANS, as well as the common global trends.
On the contrary, a surrogate-based optimization using only the small set of LES samples did not converge towards truly non-dominated designs.
Subsequently, the geometries constituting the RANS and LES Pareto fronts were compared, and highlighted major differences in performance predictions. 
Specifically, the RANS provided designs with highly deformed LE and TE compared to the LES, which triggered flow separation near the LE and was highly detrimental only in the LES.
This stemmed from the RANS boundary layers being fully turbulent, thus contributing to entropy production in much larger proportions compared to the LE flow separation, of which the relative contribution to losses was rendered negligible.
Overall, the need for scale-resolved data in the shape optimization loop was demonstrated, as well as the capacity of a MFK model to account for large deviations in objective function predictions.

To improve the present procedure, one could consider boundary layer transition turbulence models to strengthen the correlation between the two levels of fidelity.
Then, with the persistently high cost of fully integrating LES in the design loop without recourse to surrogate models (which come with their own set of disadvantages), one could envisage using LES to initially train and correct physics informed RANS models to be implemented directly in the optimization loop.
This strategy has been explored  recently in the works of Zhang \textit{et al.} \cite{Zhang2021d}, where data-driven RANS closures trained by LES are generated and employed to optimize a given geometry in a low-dimensional paramter space.
Furthermore, a possible trade-off between the detailed scale-resolving capacities of LES and its dissuasive cost for design optimization is the use of Wall-Modeled LES (WMLES) to alleviate the grid requirements in attached flows, provided the model is capable of handling transition to turbulence under pressure gradient.
Ultimately, this method would take the role of the high-fidelity function in a multi-fidelity framework.
Finally, to further reduce the cost of the present multi-fidelity optimization, one could consider performing the LES on coarse grids.
While we expect an under-resolved LES to mispredict the objective functions, we stress that the purpose of the optimization is only to identify the optimal design parameters and their correlation with the objective functions.
In this way, differences between the under-resolved LES predictions and the true blade performance may be acceptable, provided an approximate Pareto set has been identified.

\section*{Acknowledgments}
This work was granted access to the HPC resources of IDRIS and TGCC under the allocation A0162A13457 made by GENCI (Grand Equipement National de Calcul Intensif).

\clearpage
\appendix

\section{Grid convergence study}
\label{A}

\subsection{LES resolution}\label{LES resolution}
\begin{figure}[h!]
    \centering
    \includegraphics[width=0.49\textwidth]{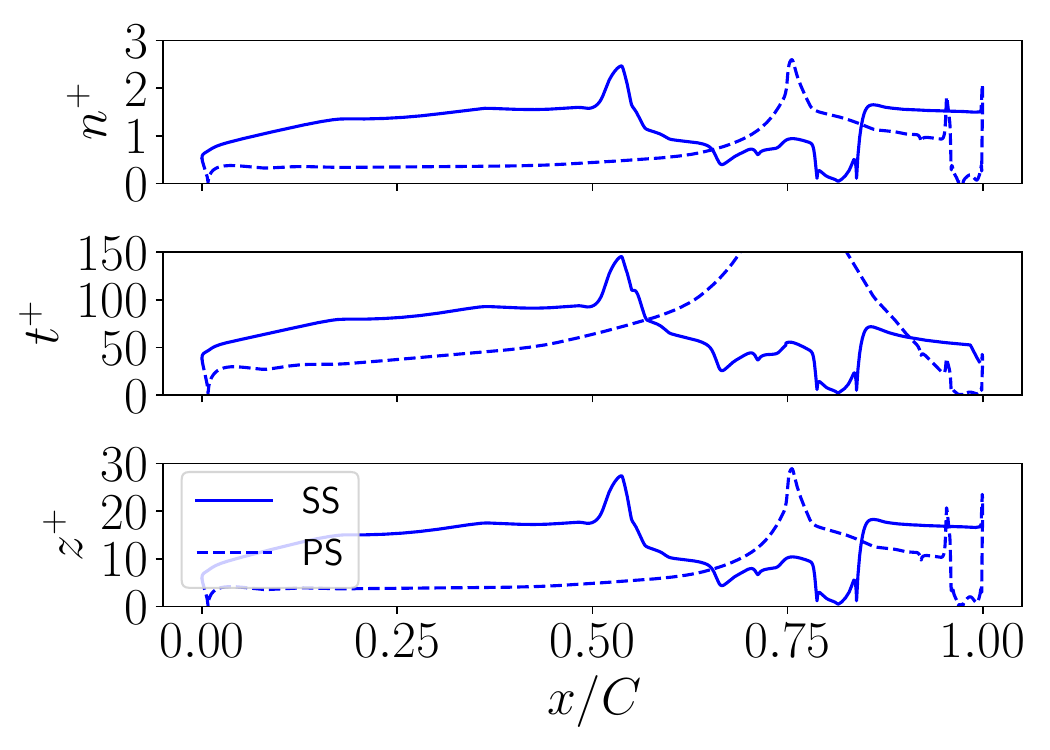}
    \caption{LES boundary layer resolution. From top to bottom: wall-normal, wall-tangential and spanwise resolutions in usual turbulent flow wall units.}
    \label{fig: LES wall resolution}
\end{figure}

We aim for the near-wall resolution achieved in \cite{Zhao2021} who investigated free-stream turbulence induced transition on a high-pressure turbine vane with mesh wall units of $n^+\approx2$, $t^+\approx30$ and $z^+\approx17$ in the wall-normal, tangential and spanwise directions, respectively, inside the turbulent boundary layer just upstream of the trailing edge (TE).
We show the evolution of these quantities on the present blade in Figure \ref{fig: LES wall resolution}, and we achieve near the TE $n^+\approx1.6$, $t^+\approx48.5$ and $z^+\approx16.8$.
While the wall-normal and spanwise resolutions correspond to that used in \cite{Zhao2021}, the present tangential resolution is somewhat lower than the target.
However, as observed in \cite{Gloerfelt2019a}, once a sufficient wall-normal resolution has been achieved, the most critical parameter becomes the spanwise resolution, which in the present study is satisfactory.
Therefore, we consider that the turbulent boundary layer is adequately resolved.

\begin{figure}[h!]
    \centering
    \includegraphics[width=0.49\textwidth]{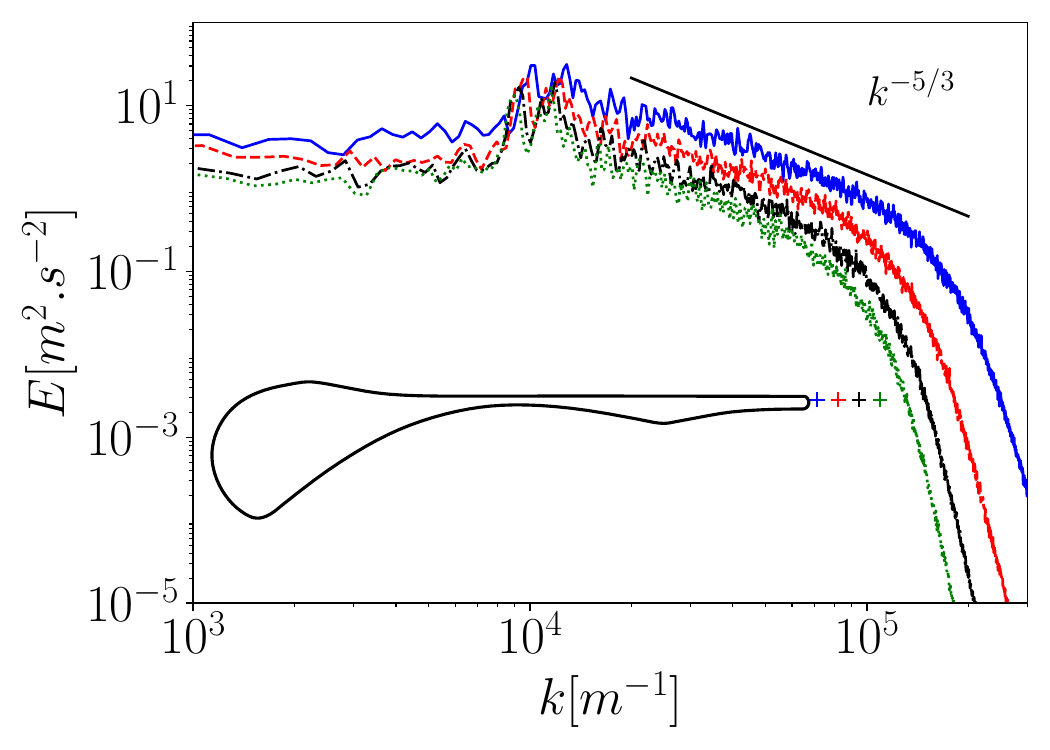}
    \includegraphics[width=0.49\textwidth]{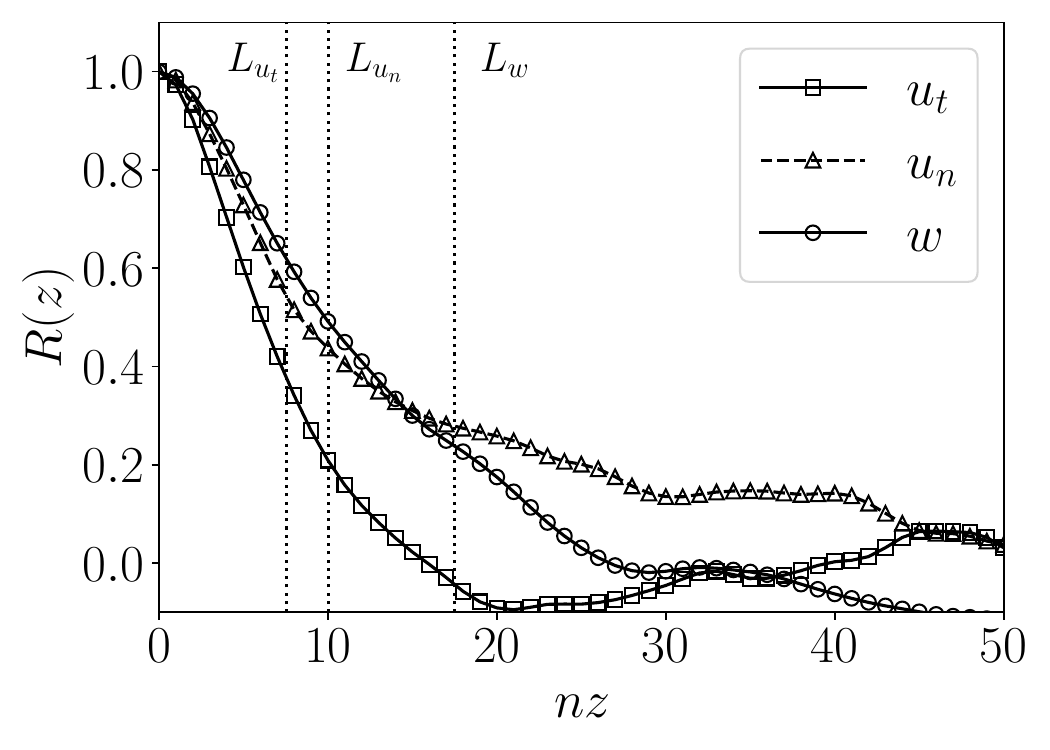}\\
    \makebox[0.49\textwidth][c]{a) Energy spectra.}
    \makebox[0.49\textwidth][c]{b) Spanwise auto-correlation.}\\
    \caption{Mesh turbulence scale resolution downstream of the trailing edge. a) Velocity components energy spectra, and b) spanwise auto-correlation functions at 3 trailing edge diameters downstream of the blade.}
    \label{fig: resolution spectrum autocorr}
\end{figure}

A major contributor to turbine flow losses is the wake developing behind the TE.
Therefore, it is paramount that the largest turbulent structures inside the wake are accurately resolved to provide sound estimates of the losses generated.
In the present work, we follow the criteria inspected in \cite{Davidson2009}.
First, a large inertial range must be observed in 1-dimensional energy spectra extracted from turbulent regions.
Here, we compute spectra from the main flow direction normal velocity component at several locations inside the turbulent wake, and observe an inertial range initially covering about 1 decade in wavenumbers, see Figure \ref{fig: resolution spectrum autocorr}.a.
Then, we compute the spanwise two-point correlations from each velocity component time signal, as well as their corresponding integral length scale in Figure \ref{fig: resolution spectrum autocorr}.
Following \cite{Davidson2009}, the required number of grid points discretizing the correlation decay until the first zero-crossing should be no-less than 6 (which corresponds to the present numerical scheme resolution limit).
In the present case, the smallest integral scale $L_{u_t}$ is discretized by at least 8 grid points, and the decay region by 16 grid points.
Therefore, the LES mesh requirements are satisfied in the present study.

\subsection{RANS grid convergence}\label{RANS grid convergence}
\begin{figure}[h!]
    \centering
    \includegraphics[width=0.49\textwidth]{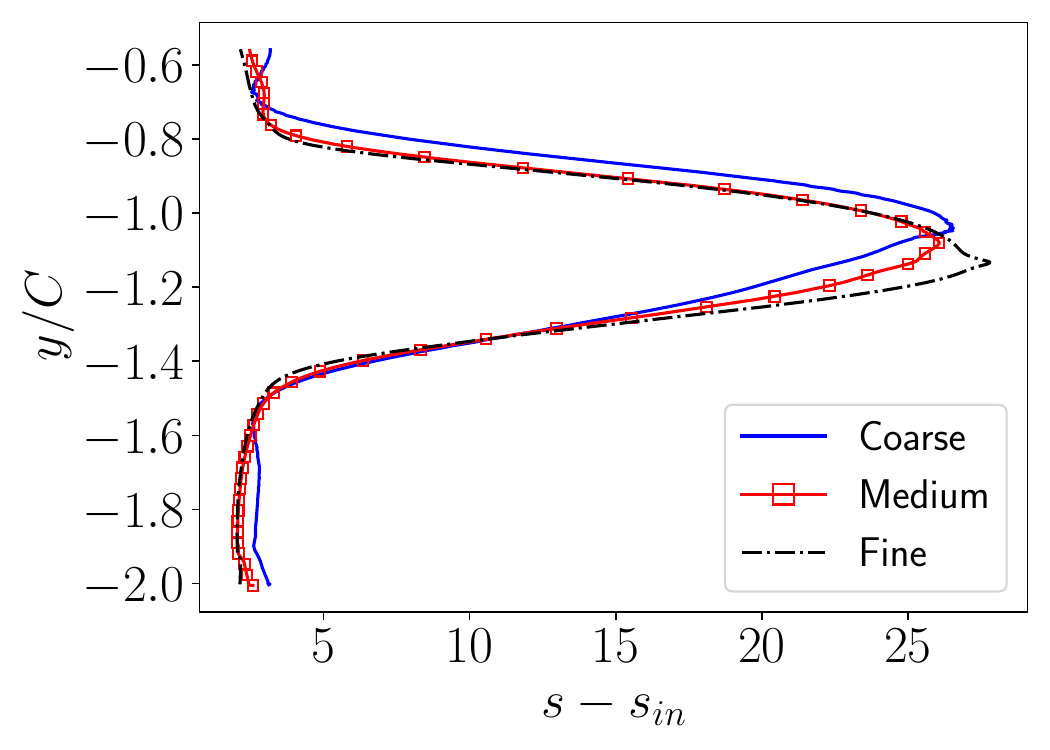}
    \includegraphics[width=0.49\textwidth]{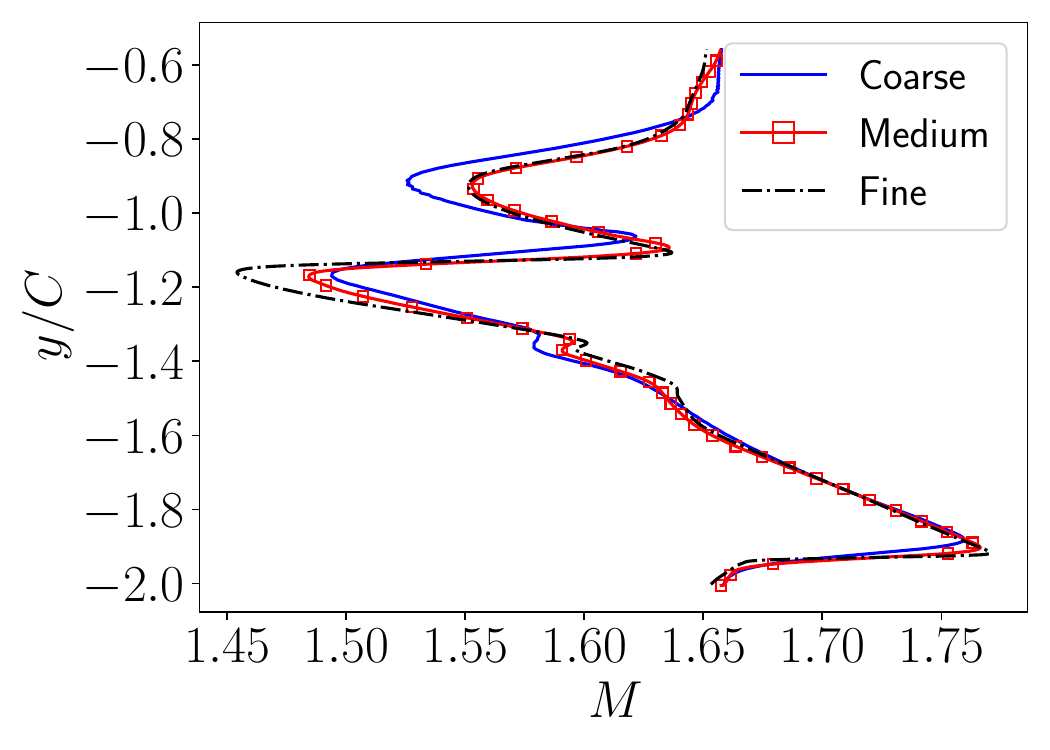}\\
    \makebox[0.49\textwidth][c]{a) Entropy deviation.}
    \makebox[0.49\textwidth][c]{b) Mach number.}\\
    \caption{Evolution of RANS entropy and Mach number profiles, at $x/C=1.5$ along pitch, with mesh refinement.}
    \label{fig: RANS mesh convergence}
\end{figure}

We assess the convergence of the RANS solution with mesh refinement by comparing entropy deviation and Mach number profiles extracted behind the blade cascade, where the optimization objective functions are computed.
Three grids are designed, with 400, 800 and 1\,800 points discretizing the blade, leading to a total of 35\,625, 142\,500 and 840\,000 points, respectively.
The extracted profiles are provided in Figure \ref{fig: RANS mesh convergence}.
Besides the slight change in predicted maximum entropy and minimum Mach number inside the wake, fair agreement of the solutions on the medium and fine grids is obtained.
Thus, we retain the medium grid as reference for the RANS computations.

\section{Free Form Deformation}
\label{B}
\begin{figure}[h!]
    \centering
    \includegraphics[width=0.5\textwidth]{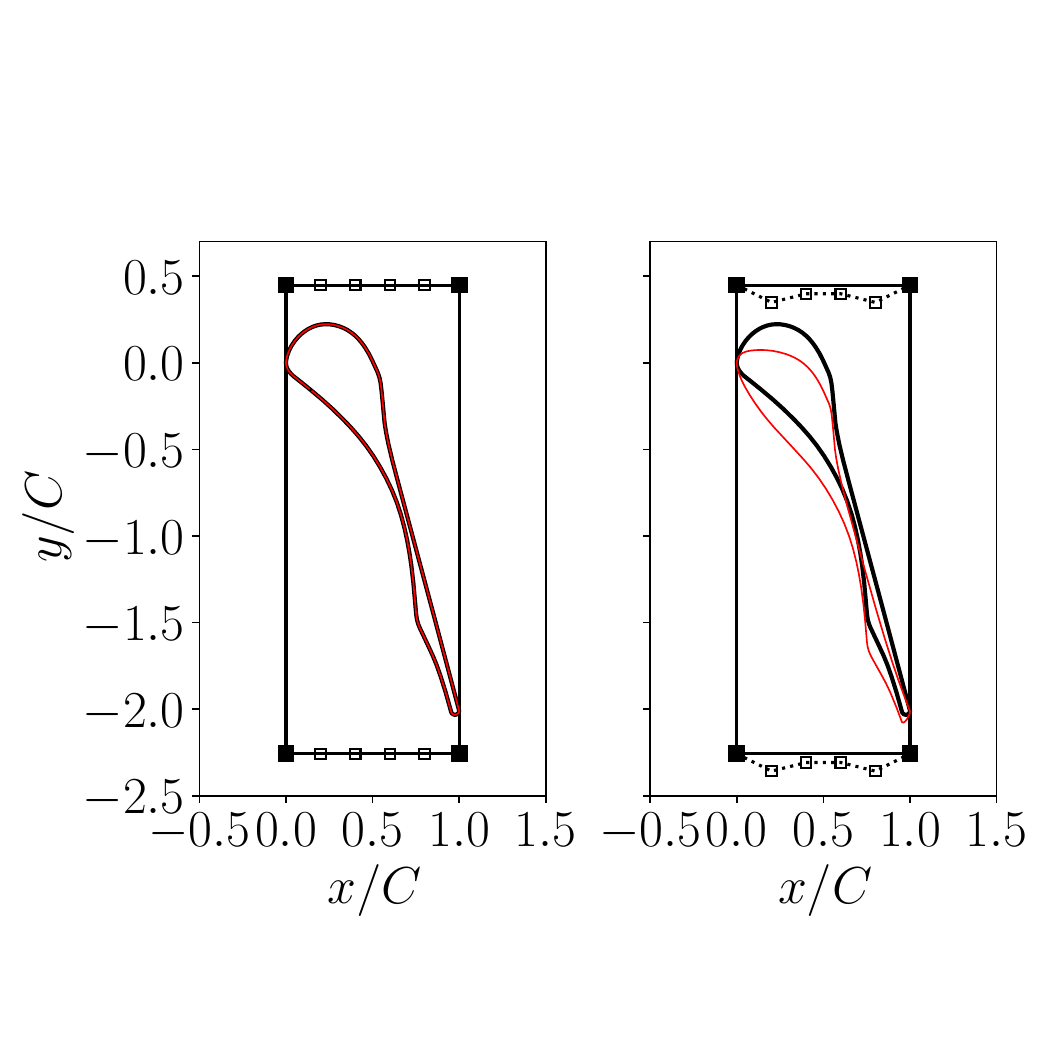}
    \caption{Example of FFD parametrization applied to the present turbine blade.}
    \label{fig: example FFD}
\end{figure}

To parametrize the blade with FFD, we first create a lattice surrounding the geometry made of 12 control points, see Figure \ref{fig: example FFD}. 
Then, we fix the 4 outermost corners (the black filled squares) and leave the remaining 8 free to move (the empty squares).
The example shows the effect of displacing the 8 control points downwards.





\bibliographystyle{elsarticle-num} 
\bibliography{Biblio_these}

\end{document}